\documentclass[journal]{IEEEtran}

\ifdefined\pdfobjcompresslevel\pdfobjcompresslevel=0\fi
\ifdefined\pdfminorversion\pdfminorversion=7\fi

\usepackage{amsmath,amsfonts}
\usepackage{algorithmic}
\usepackage{algorithm}
\usepackage{array}
\usepackage{xcolor}
\usepackage[caption=false,font=footnotesize]{subfig}
\usepackage{textcomp}
\usepackage{stfloats}
\usepackage{url}
\usepackage{verbatim}
\usepackage{graphicx}
\usepackage{booktabs}
\usepackage{amssymb}
\usepackage{makecell}
\usepackage{tabularx}
\usepackage[style=ieee,backend=biber]{biblatex}

\usepackage{enumitem}
\usepackage{setspace}
\usepackage{bytefield}
\usepackage{nth}
\usepackage{multirow}
\usepackage{tikz}
\usetikzlibrary{calc, shapes.geometric}
\usepackage{standalone}
\usepackage{cleveref}
\usepackage{float}
\crefname{section}{§}{§§}
\Crefname{section}{§}{§§}

\definecolor{lightcyan}{rgb}{0.84,1,1}
\definecolor{lightgreen}{rgb}{0.64,1,0.71}
\definecolor{lightred}{rgb}{1,0.7,0.71}
\definecolor{white}{rgb}{1.0,1.0,1.0}

\usepackage[nowarn,acronyms,nonumberlist,nopostdot,nomain,nogroupskip]{glossaries}

\newlist{SubItemList}{itemize}{1}
\setlist[SubItemList]{label={$-$}}

\let\OldItem\item
\newcommand{\SubItemStart}[1]{%
	\let\item\SubItemEnd
	\begin{SubItemList}[resume]%
		\OldItem #1%
		}
		\newcommand{\SubItemMiddle}[1]{%
			\OldItem #1%
		}
		\newcommand{\SubItemEnd}[1]{%
	\end{SubItemList}%
	\let\item\OldItem
	\item #1%
}
\newcommand*{\SubItem}[1]{%
	\let\SubItem\SubItemMiddle%
	\SubItemStart{#1}%
}%

\newacronym{3gpp}{3GPP}{3rd Generation Partnership Project}
\newacronym{5g}{5G}{Fifth-Generation Mobile Network}
\newacronym{6g}{6G}{Sixth-Generation Mobile Network}
\newacronym{5gc}{5GC}{5G Core}
\newacronym{aoa}{AoA}{Angle of Arrival}
\newacronym{aod}{AoD}{Angle of Departure}
\newacronym{api}{API}{Application Programming Interface}
\newacronym{achem}{ACHEM}{A Channel Emulator}
\newacronym{ber}{BER}{Bit Error Rate}
\newacronym{cfo}{CFO}{Carrier Frequency Offset}
\newacronym{chem}{CHEM}{Channel Emulator}
\newacronym{dac}{DAC}{Digital-to-Analog Converter}
\newacronym{darpa}{DARPA}{Defense Advanced Research Projects Agency}
\newacronym{dt}{DT}{Digital Twin}
\newacronym{dl}{DL}{Downlink}
\newacronym{enb}{eNB}{Evolved Node B}
\newacronym{epc}{EPC}{Evolved Packet Core}
\newacronym{fifo}{FIFO}{First In First Out}
\newacronym{fir}{FIR}{Finite Impulse Response}
\newacronym{fpga}{FPGA}{Field Programmable Gate Array}
\newacronym{gnb}{gNB}{Next-Generation Node B}
\newacronym{gpsdo}{GPSDO}{GPS Disciplined Oscillator}
\newacronym{itu-t}{ITU-T}{ITU Telecommunication Standardization Sector}
\newacronym{kpi}{KPI}{Key Performance Indicator}
\newacronym{lte}{LTE}{Long-Term Evolution}
\newacronym{otw}{OTW}{Over-the-Wire}
\newacronym{oran}{O-RAN}{Open Radio Access Network}
\newacronym{pci}{PCI}{Physical Cell Identity}
\newacronym{prb}{PRB}{Physical Resource Block}
\newacronym{rf}{RF}{Radio Frequency}
\newacronym{rsrp}{RSRP}{Reference Signal Received Power}
\newacronym{sdr}{SDR}{Software Defined Radio}
\newacronym{snr}{SNR}{Signal-to-Noise Ratio}
\newacronym{simd}{SIMD}{Single Instruction Multiple Data}
\newacronym{hitl}{HITL}{Hardware-in-the-Loop}
\newacronym{sitl}{SITL}{Software-in-the-Loop}
\newacronym{uhd}{UHD}{USRP Hardware Driver}
\newacronym{vusrp}{V-USRP}{Virtual Universal Software Radio Peripheral}
\newacronym{uav}{UAV}{Uncrewed Aerial Vehicle}
\newacronym{ue}{UE}{User Equipment}
\newacronym{ul}{UL}{Uplink}
\newacronym{ugv}{UGV}{Uncrewed Ground Vehicle}
\newacronym{usrp}{USRP}{Universal Software Radio Peripheral}
\newacronym{uuid}{UUID}{Universally Unique Identifier}
\newacronym{aerpaw}{AERPAW}{Aerial Experimentation and Research Platform for Advanced Wireless}
\newacronym{agc}{AGC}{Automatic Gain Control}
\newacronym{cdf}{CDF}{Cumulative Distribution Function}
\newacronym{cir}{CIR}{Channel Impulse Response}
\newacronym{dsp}{DSP}{Digital Signal Processing}
\newacronym{fdd}{FDD}{Frequency-Division Duplex}
\newacronym{nr}{NR}{New Radio}
\newacronym{tdd}{TDD}{Time-Division Duplex}

\begin{document}

\author{{An\i l G\"urses,~\IEEEmembership{Student Member,~IEEE}, Mihail L. Sichitiu,~\IEEEmembership{Member,~IEEE}}
	\thanks{This work was funded in part by NSF award CNS-1939334. The software is available publicly~\cite{achem}.}}

\title{ACHEM: A Real-Time Digital Twin Framework with Channel and Radio Emulation}

\markboth{ACHEM: A Real-Time Digital Twin Framework with Channel and Radio Emulation}%
{}


\maketitle

\begin{abstract}
	Digital twins are becoming an important tool for designing, developing, testing, and optimizing next-generation wireless communication systems.
	Over the past decade, system softwarization has become a reality, and wireless communication systems are no exception.
	Software-Defined Radios (SDRs), in general, and Universal Software Radio Peripherals (USRPs), in particular, are often used for prototyping and testing advanced wireless systems.
	Unfortunately, there is currently no end-to-end, software-based, general-purpose testing environment for SDR-based systems: developers often rely on benchtop setups or even small testbeds, but those are costly and cumbersome to build.
	At the other end of the spectrum, simulations often rely on simplified channel/radio models and typically do not execute full-stack production code, which can increase development effort and reduce fidelity.
	In this paper, we propose ACHEM (A Channel Emulator), the first software-based, end-to-end wireless channel emulation environment and toolset for communication systems based on SDRs, specifically USRPs.
	With the proposed emulator and toolkit, any USRP-based system can be fully emulated at the I/Q level in a pure digital environment without requiring specialized hardware (e.g., vehicles, USRPs, FPGAs, or GPUs).
	The proposed emulator supports multiple transmitters and receivers, MIMO communications, multiple frequencies, heterogeneous sampling rates, real-time node mobility through vehicle emulation, antenna radiation patterns, and various channel models.
	ACHEM facilitates wireless digital twin development and deployment.
	ACHEM is validated with several popular open-source USRP-based wireless communication applications, including GNU Radio, srsRAN 4G/5G, and OpenAirInterface.
\end{abstract}


\section{Introduction}

\glsresetall

Throughout the last decade, \glspl*{sdr} have received significant attention due to their ease of use for wireless communication systems, radars, spectrum sensing, and even visible light communications \cite{usrp_survey,usrp_vlc}.
\glspl*{sdr} are used for the development and testing of wireless systems, and they provide great flexibility in terms of supporting different applications.
The \gls*{usrp} is an early and widely adopted \gls*{sdr} hardware in this field \cite{usrp_spectrum_artc}.
Its open-source nature and robust support from the community have made it a widely adopted solution for many researchers, hobbyists, and professionals.
With open-source tools available today \cite{oai_5g,srslte}, it is possible to run end-to-end \gls*{lte} and \gls*{5g} stacks with consumer-grade computers and inexpensive \glspl*{sdr}, such as \glspl*{usrp}.
This accessibility significantly lowers the barriers to entry for wireless communication innovation and experimentation.

\begin{figure}[ht]
	\centering
	\includegraphics[width=0.99\columnwidth]{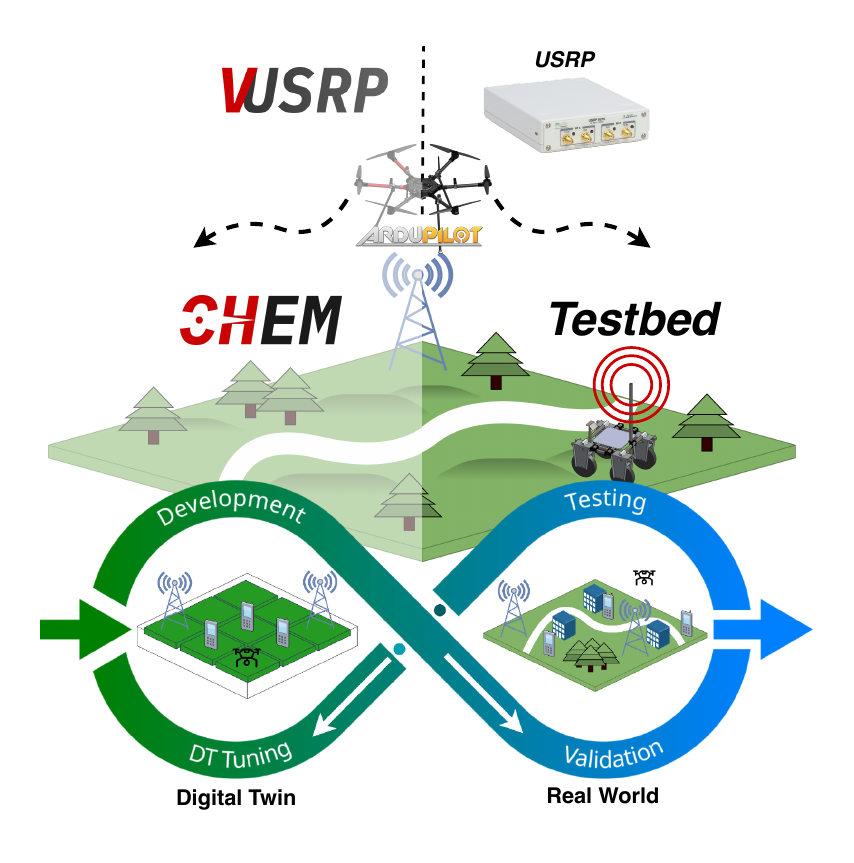}
	\caption{The digital twin system that is proposed in this work: a digital twin with virtualized components (left) replicates a real communication network (right).}
	\label{fig:dt_overview}
\end{figure}

While \glspl*{sdr} are practical and versatile tools for wireless communication system development and testing, they introduce additional complexity and overhead.
This can be particularly challenging for researchers who are unfamiliar with the nuances of \glspl*{sdr} or lack access to the necessary resources.
As a result, simulations are often the primary and essential tool to assess the potential merits of new wireless systems.
Simulations offer valuable insights into the behavior of wireless systems and help researchers evaluate the performance of new systems under various scenarios.
However, simulations often fall short of replicating the complexities of the system and real-world dynamics.
Wireless communication performance depends on many factors, such as multipath propagation, mobility, antenna patterns, and clock synchronization.
Simulations often rely on  {\em statistical channel models} and on assumptions that gloss over important channel impairments.
Additionally, most simulators rely on abstractions of systems, where only specific components of the system are considered.
Emulation, on the other hand, provides a more realistic environment for replicating a real system while using production code without modifications, offering a more accurate representation of system behavior under various scenarios.

The increasing availability of aforementioned tools, together with advances in computational capabilities, has facilitated the development of digital twins for wireless communication systems.
Digital twins can be broadly classified into two categories: offline digital twins and online digital twins.
An offline digital twin, as illustrated in Fig. \ref{fig:dt_overview}, is a digital replica of a communication system used to evaluate, debug, and optimize behavior under controlled, repeatable conditions.
Depending on the use case, a digital twin can be purely digital or incorporate hardware components in the loop~\cite{qian2022digital}.
An online digital twin extends an offline digital twin by incorporating real-time system data to enable continous monitoring, optimization, and decision making.

By leveraging digital twins, researchers can evaluate and test wireless communication systems under controlled, repeatable conditions, improving reproducibility while reducing the cost and effort of experimentation.
Although digital twins can be realized using simulations, abstractions of radio interfaces and protocol implementations can limit their ability to capture the behavior of deployed software.
Emulation, on the other hand, allows production software to interact with emulated radio interfaces and channels, preserving implementation-specific behavior that may be abstracted away in simulations.
In this work, we focus on emulation as a means of supporting offline digital twins for wireless communication systems based on \glspl*{usrp} and \glspl*{sdr}.
The overall system architecture of the proposed digital twin framework is illustrated in Fig. \ref{fig:dt_overview}.

Wireless sytems can be emulated at least at three different levels: 1) at packet level, where the unit of information exchanged is a network packet \cite{ns3}, 2) at the I/Q level, where the unit is a baseband I/Q sample, and 3) at the \gls*{rf} level, where the RF channel is captured in the up-converted frequency band.
RF emulators are constructed using real wireless hardware connected to the emulation hardware via RF cables \cite{peter_emu,hil_emu,propsim}.
Colosseum \cite{colosseum} is a prominent example of RF-level emulation; to the best of our knowledge, it is the only RF emulator that can support up to 256 nodes.
Recent work has also demonstrated Colosseum as a digital twin of an over-the-air wireless testbed~\cite{colosseum_dt}.
Although RF emulators provide extensive realism, they come with a considerable build cost and quickly become outdated due to advancements in radio hardware and computing technologies.
At this stage, software-based emulators represent a suitable middle ground for realizing digital twins.
Recently, NVIDIA and a few other companies have started to provide digital twin solutions for wireless communication applications, but they are not tailored for portability to production environments as envisioned in this work \cite{nvidia_aodt}.

In this work, we present \gls*{achem}, an open-source, real-time digital twin framework for \gls*{usrp}-based wireless systems.
\gls*{achem} emulates wireless propagation at the I/Q level, enabling various channel impairments, heterogeneous configurations, and node mobility in a fully software-based environment.
To enable compatibility with existing application software, \gls*{achem} includes a \gls*{vusrp} that allows unmodified \gls*{uhd}-based applications to run within the emulation environment.
The \gls*{vusrp} replicates the functionality of real USRP hardware, including multiple RX/TX ports, various sampling rates, timed commands, and flexible frequencies.
While this paper validates \gls*{achem} on the \gls*{aerpaw} testbed, the framework is general-purpose and applicable to other \gls*{usrp}-based experimentation environments.
The focus of this work is to provide a reusable radio and channel emulation infrastructure for constructing digital twins and facilitating the transfer of experiments between emulated and physical environments.

In this paper, we make the following contributions:
\begin{itemize}
	\item We introduce a radio emulator \gls*{vusrp} that emulates the functionality of real \glspl*{usrp};
	\item We introduce \gls*{chem}, which:
	      \SubItem{coordinates wireless propagation between multiple nodes with varying frequencies and sampling rates;}
	      \SubItem{emulates wireless channel impairments and antenna effects;}
	      \SubItem{supports mobile nodes by integrating with existing \gls*{uav}/\gls*{ugv} emulators;}
	\item We validate \gls*{chem} and \gls*{vusrp} using open-source \gls*{lte} and \gls*{5g} stacks, and compare the results from \gls*{achem} with those from a real testbed.
\end{itemize}

The rest of the paper is organized as follows: Section~II provides an overview of a real-world USRP-based wireless communication scenario and its representation in \gls*{achem}.
Section~III details the design of the \gls*{vusrp}, \gls*{chem}, and overall framework, including the channel processing flow.
Section~IV validates \gls*{achem} using \gls*{lte} and \gls*{5g} communication stacks and compares the results with real-world experiments.
Section~V highlights the key performance metrics of \gls*{achem}.
Section~VI provides an overview of related work and compares \gls*{achem} with existing solutions.
Finally, Section~VII concludes the paper.

\begin{figure*}[!htp]
	\centering
	\subfloat[An example end-to-end system setup with a UAV carrying a portable node and a fixed node, each with a USRP device]{%
		\includegraphics[height=0.8\columnwidth]{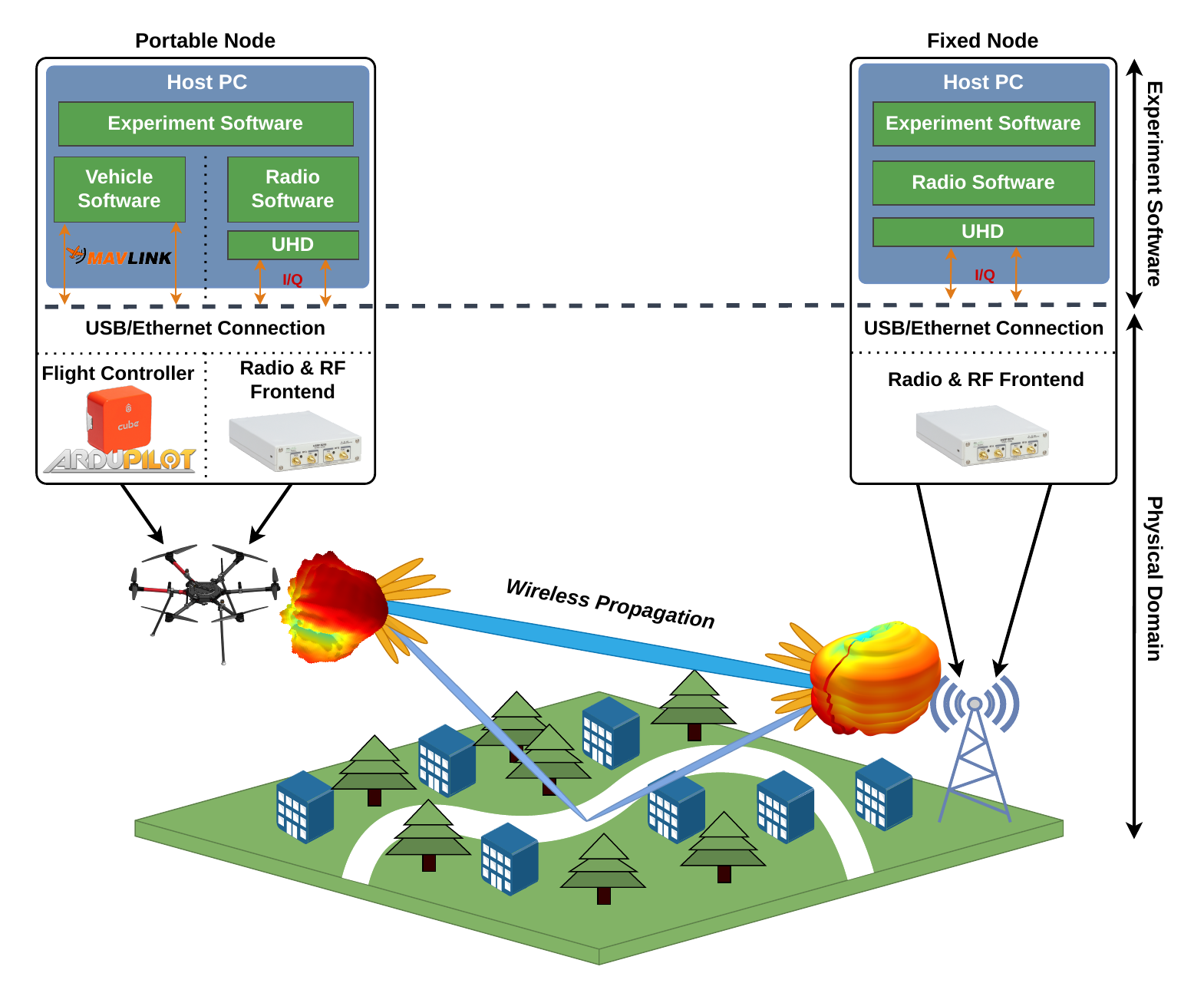}
		\label{subfig:exp_system}
	}
	\hfill
	\subfloat[Equivalent setup with V-USRP and CHEM, with a portable node and a fixed node.]{%
		\includegraphics[height=0.8\columnwidth]{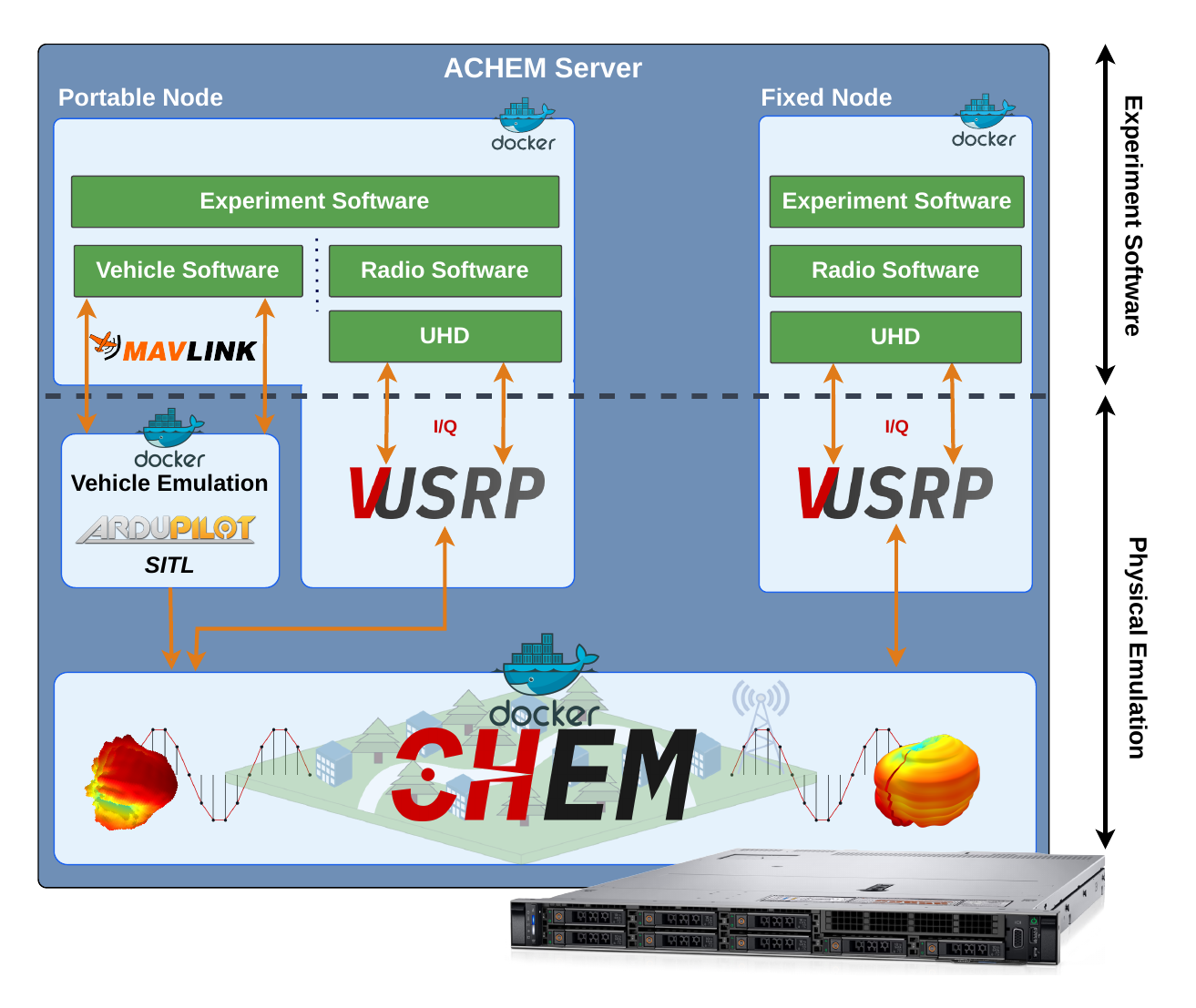}
		\label{subfig:achem_system}
	}
	\caption{A communication system with USRPs in a real testbed and in the ACHEM framework.}
	\label{fig:system}
\end{figure*}
\section{Overall System Design}

\glsunset{achem}
In this section, we present a sample scenario to illustrate how a USRP-based wireless system can be implemented and deployed first with real hardware and then in our digital twin environment.

\subsection{Sample Scenario}

Fig.~\ref{subfig:exp_system} illustrates a sample end-to-end \gls*{usrp}-based experiment setup with two \glspl*{usrp} and two host PCs, where one node is mobile and the other is fixed.
Each of the experiment nodes consists of a host PC and a \gls*{usrp} radio, and the first node is carried by a \gls*{uav} while the second one is fixed on a tower.
Nodes run experiment software (e.g., srsRAN \cite{srslte} and GNU Radio \cite{gnuradio}) that communicates with the \glspl*{usrp} through the \gls*{uhd} driver; the vehicle for node one is controlled by dedicated vehicle-control software.
Vehicle software is responsible for communicating with vehicle control hardware and providing autonomous operation capabilities.
The setup on the physical testbed requires several people, and it is not portable to other locations without significant effort and cost.

This sample scenario can be seamlessly transitioned into a digital twin by using the \gls*{achem} environment without modifying the experiment software.
Portability of \gls*{usrp} application software is achieved by introducing \glspl*{vusrp} (a component of \gls*{achem}), which allows existing \gls*{usrp} software to run unaltered within the \gls*{achem} environment and enables a seamless transition between the real world and the digital twin.
In \gls*{achem}, the experiment software from all nodes can run together on the same computer, either on bare metal or with virtualization technology (e.g., Docker), for increased flexibility and isolation.
Although \gls*{achem} can run on bare metal, containers or virtual machines are simpler for logically separating processes across different experiment nodes.
In Section~IV, we employ Docker containers to provide this logical separation while minimizing virtualization overhead.
Node mobility is supported by a vehicle emulator; in this study, we use ArduPilot \gls*{sitl}.
\gls*{achem} coordinates both the \glspl*{vusrp} and the \gls*{sitl} instances (one per vehicle) corresponding to each physical node.
In addition to the logical separation, the containerized environment facilitates the transition of \gls*{achem}-based experiments into a real testbed thanks to the portability of Docker containers.

\subsection{ACHEM Overview}\label{sec:emu_overview}

\begin{figure*}[ht]
	\centering
	\includegraphics[height=0.525\textwidth]{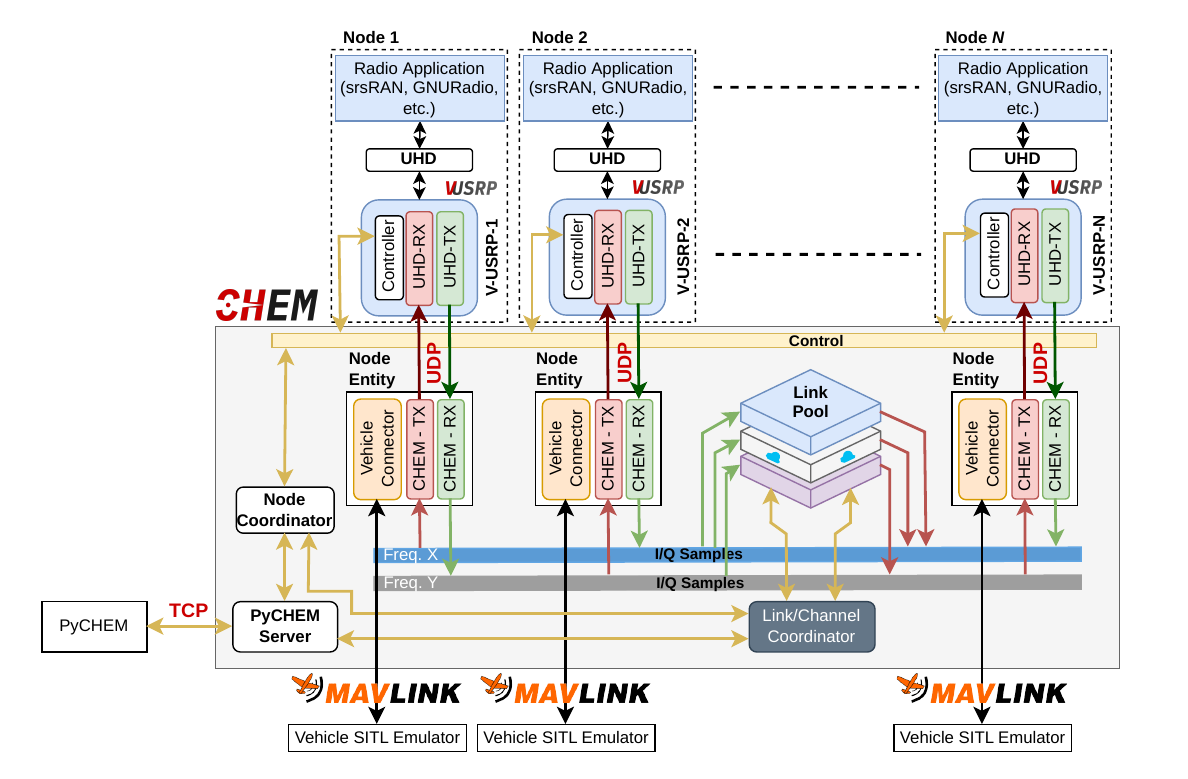}
	\caption{Overall \gls*{achem} architecture and \gls*{vusrp}/\gls*{chem} integration. N mobile nodes transmit and receive concurrently on two frequencies in this example.}
	\label{fig:achem_vusrp_arch}
\end{figure*}

\gls*{achem} is a digital twin framework that emulates wireless channels and radios with two unique components: \gls*{chem} and \gls*{vusrp}.
\gls*{chem} is the digital counterpart of real-world wireless propagation and the \gls*{vusrp} corresponds to the \gls*{usrp} device.
\gls*{chem} emulates the wireless channel at the I/Q level in real time, making it protocol-agnostic and allowing it to operate without application-software changes.
Additionally, \gls*{achem} provides an experiment-management tool called PyCHEM for configuring the emulation environment, including propagation models, antenna radiation patterns, and channel coefficients.

The overall architecture and interconnections of the \gls*{chem} and \gls*{vusrp} are shown in Fig.~\ref{fig:achem_vusrp_arch}.
\gls*{chem} provides orchestration for the nodes, wireless channels, and node mobility.
A single \gls*{chem} instance is used for an arbitrary number of \glspl*{vusrp}.
This configuration allows nodes within an environment to operate synchronously at the same center frequency or at different center frequencies.
The ability to support multiple center frequencies enables the emulation of complex wireless communication environments, where devices may be required to transmit and receive on distinct channels, including \gls*{tdd} and \gls*{fdd} configurations.

Similar to the way a \gls*{usrp} transfers signals between radio application software and the real world, the \gls*{vusrp} couples the application software and \gls*{chem}.
Each \gls*{vusrp} has three components working in parallel that communicate separately with \gls*{chem}.
\gls*{chem} receives the I/Q samples through UDP sockets from each \gls*{vusrp} and processes the I/Q samples based on channel information for each node pair.
Channel information is acquired from multiple sources, including path distance from the Vehicle Connector, path-loss model configuration from the Channel Coordinator, and antenna patterns from the Node Coordinator.
Finally, the processed I/Q samples are transferred to the receiving \glspl*{vusrp} through UDP sockets.

\gls*{chem} comprises five main components: Node Entity, Node Coordinator, Link/Channel Coordinator, PyCHEM Connector, and Link Pool.
A Node Entity communicates with each node's \gls*{vusrp} and controls node-specific settings, such as antenna radiation pattern and path loss model.
All the node configurations and related communications are handled by the Node Coordinator on the \gls*{chem} side, and the corresponding Controller on the \gls*{vusrp} side.
The Link/Channel Coordinator is responsible for creating and destroying virtual wireless channels/links for frequencies where at least a transmitting and a receiving \glspl*{vusrp} are tuned to.
The created channels are allocated in the link pool and the Link/Channel Coordinator chooses the respective channel for each stream based on the center frequency of the stream.
Lastly, the PyCHEM Connector handles the communications with the PyCHEM library/user interface and forwards the requested changes to both the Node Coordinator and Link/Channel Coordinator.

\gls*{chem} is designed to support mobility updates from MAVLink sources, such as the ArduPilot~\cite{ardupilot} and PX4~\cite{px4} \gls*{sitl} vehicle emulators.
ArduPilot is a popular open-source firmware for \glspl*{uav}, \glspl*{ugv}, and many other types of vehicles \cite{ardupilot}.
Using the \gls*{sitl} feature of ArduPilot, any compatible vehicle can be accurately emulated at the software level.
The communication between \gls*{chem} and the vehicle emulator is handled through TCP sockets and MAVLink messages \cite{mavlink}.
MAVLink is a popular messaging protocol for \glspl*{uav} and \glspl*{ugv}.

\section{System Implementation}
The \gls*{achem} emulation system consists of two main components: \gls*{vusrp} and \gls*{chem}.
In this section, we first explain the implementation of \gls*{vusrp} and then \gls*{chem}.
We then explain how the \glspl*{vusrp} interact with \gls*{chem}.

\subsection{Virtual USRP~(V-USRP)}

\begin{figure*}[ht]
	\centering
	\subfloat[Experimental setup with a real \gls*{usrp} device.]{%
		\includegraphics[width=0.99\columnwidth,trim={0 0 0 0.5cm},clip]{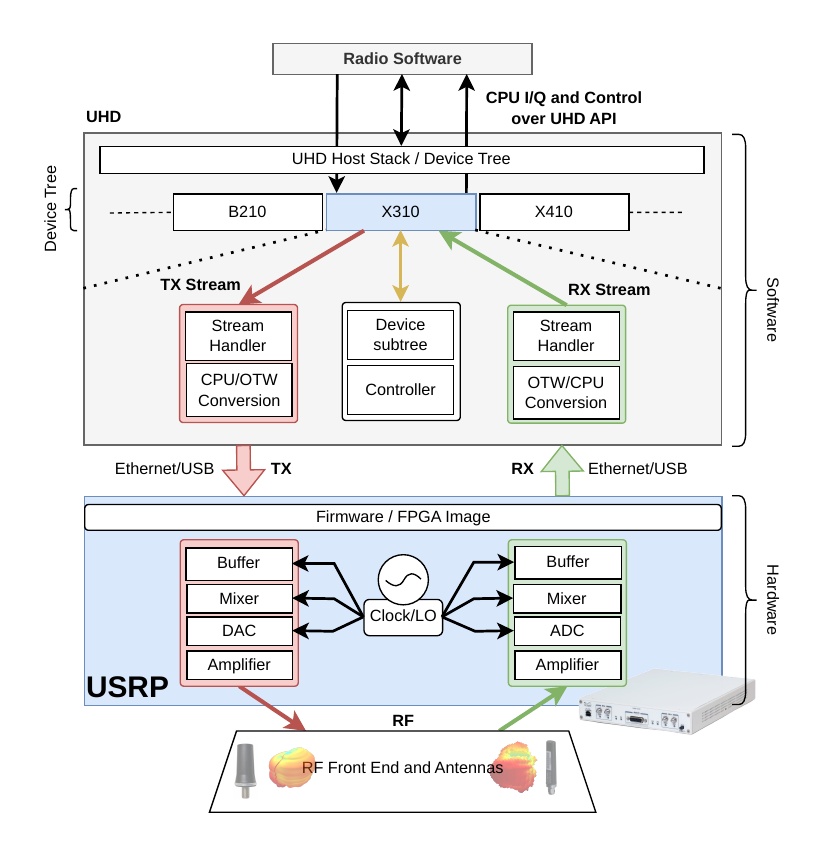}
		\label{subfig:usrp_system}
	}
	\hfill
	\subfloat[Experimental setup with \gls*{vusrp} and \gls*{chem}.]{%
		\includegraphics[width=0.99\columnwidth,trim={0 0 0 0.5cm},clip]{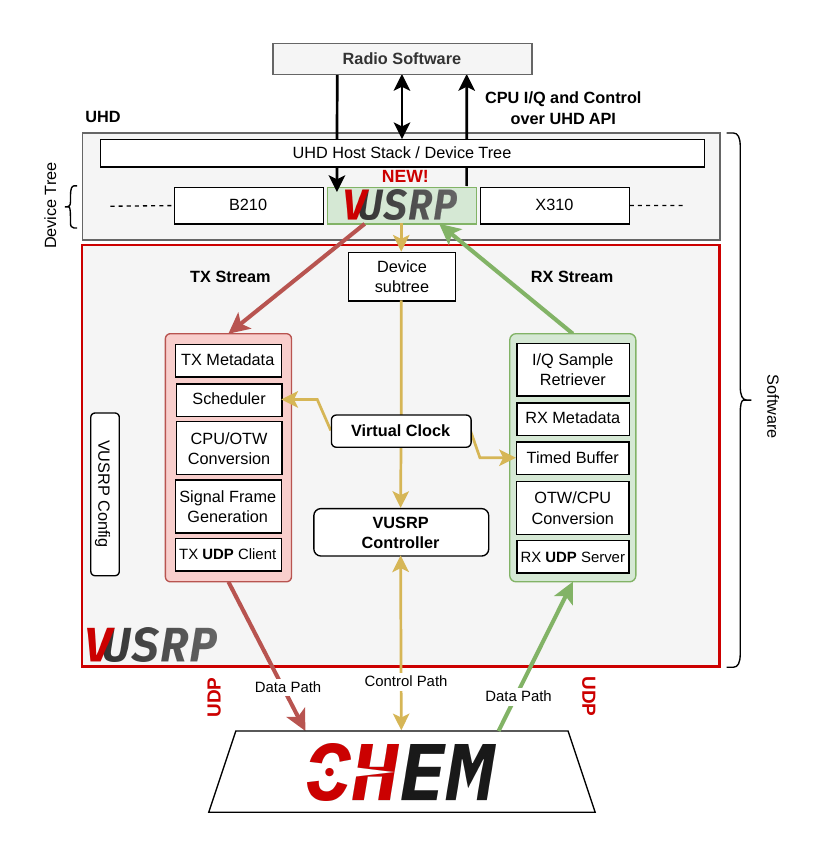}
		\label{subfig:vusrp_system}
	}
	\caption{Comparison of (a) X310 USRP and (b) V-USRP.}
	\label{fig:vusrp_vs_usrp}
\end{figure*}

For this work, we forked the open-source \gls*{uhd} driver \cite{uhd} and developed a new device type called \gls*{vusrp} within the driver.
The \gls*{vusrp} emulates real \gls*{usrp} hardware for \gls*{uhd}, allowing unmodified applications that use \gls*{uhd} to run the same way as on \gls*{usrp} hardware.
A typical hardware \gls*{usrp} communicates with the host PC through Ethernet or USB, as shown in Fig.~\ref{subfig:usrp_system}.
The \gls*{vusrp} uses network interfaces, specifically UDP sockets, for the same purpose as shown in Fig.~\ref{subfig:vusrp_system}.
The communication between real \gls*{usrp} and the host PC is replicated by the \gls*{vusrp}.

In a real testbed setup, a single PC can control multiple USRPs, and each USRP may have multiple RX and TX streams (phase coherent or not).
Our emulation system supports the same features (i.e., multiple V-USRPs per node and, for each V-USRP, multiple MIMO RX and TX streams); however, for clarity, we present scenarios with a single V-USRP per host and one TX and one RX stream per V-USRP.

The application software may create multiple TX and RX streams during initialization depending on the application requirements.
A stream includes properties such as sampling rate, center frequency, data format, transmit/receive type, and special commands.
A real \gls*{usrp} uses stream properties for configuring the RF components and uses in-hardware buffering for continuous delivery of I/Q samples to and from the application software.
In the \gls*{vusrp}, this is achieved by sending the stream configuration information to \gls*{chem}, which instantiates a Node Entity for each joining \gls*{vusrp}.
The \gls*{vusrp} instantiates a TX UDP client for transmitting and a RX UDP server for receiving I/Q samples to/from \gls*{chem}.

In a real \gls*{usrp}, the transmitted I/Q samples from the application software to the \gls*{uhd} are represented in a so-called ``CPU format'' in the \gls*{api} call to the \gls*{uhd}, which converts them to an \gls*{otw} format to send them to the \gls*{usrp}.
At reception, the reverse conversion occurs.
To preserve the emulation fidelity, the \gls*{vusrp} also converts the I/Q samples between the two formats, in both directions.

\begin{description}[style=unboxed,leftmargin=0cm]
	\item[Signal frame:] \label{subsec:signal}
	      Similar to a real \gls*{usrp}, the \gls*{vusrp} has two logical paths: the data path and the control path.
	      On the data path, the \gls*{vusrp} transfers I/Q samples, while on the control path it transfers configuration-related information such as timing, sampling rate, and center frequency.
	      The data exchanged between the \gls*{vusrp} and \gls*{chem} is encapsulated in a {\em signal frame}, shown in Fig.~\ref{fig:signal_frame}, that includes the I/Q samples as well as several control fields associated with those I/Q samples.
	      The first field carries the emulated transmit time used to place the frame on the radio timeline.
	      For scheduled transmissions, this value is the transmit time requested through the \gls*{uhd} metadata; for unscheduled transmissions, the \gls*{vusrp} assigns the next available transmission instant.
	      This field is used to order I/Q samples at the receiver.
	      The second field carries the actual transmission time, obtained from the emulation host clock, and is used for computing propagation delay.
	      The third and fourth fields carry the number of samples and the number of channels, respectively.
	      I/Q samples fill the remaining frame space.
	      Samples from MIMO channels are multiplexed in the same frame to maintain temporal alignment of I/Q samples.
	      Signal frame size is determined by the sampling rate and number of channels, and it is configured by the application software through the \gls*{uhd} \gls*{api}.
	      Before transmission, each signal frame is fragmented into multiple UDP datagrams sized according to the maximum transmission unit (MTU) of the network interface.
	      The custom UDP fragmentation and reassembly process is implemented to avoid the overhead of IP fragmentation and a complete drop of an entire frame, while maintaining low latency in the system.

	      \begin{figure}[t]
		      \centering
		      \includegraphics[width=\columnwidth]{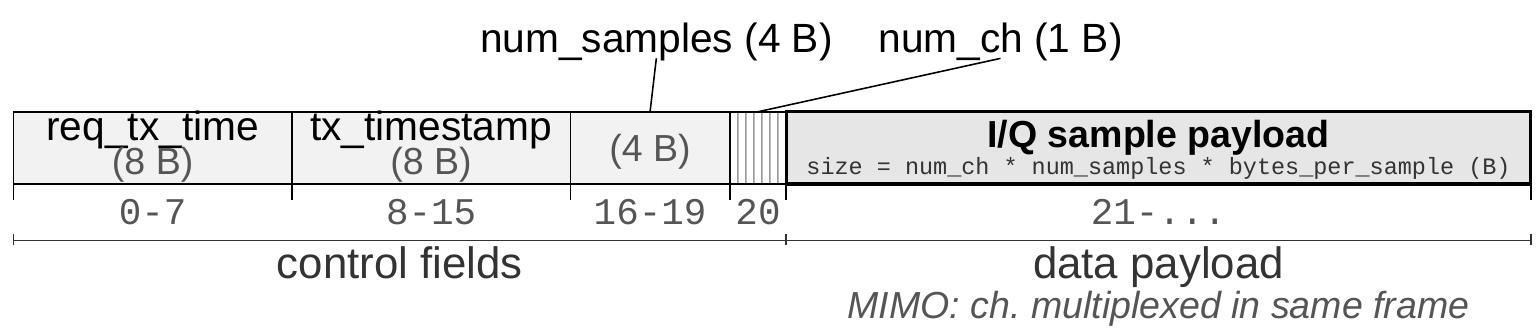}
		      \caption{Signal frame structure for data exchange between V-USRP and CHEM.}
		      \label{fig:signal_frame}
	      \end{figure}

	\item[Transmitting:]
	      The \gls*{vusrp} transmitter employs a TX UDP client to send I/Q samples, encapsulated as signal frames, from the \gls*{vusrp} to \gls*{chem}.
	      A UDP client is initialized when the application software creates a TX stream.
	      To avoid unnecessary processing overhead, \gls*{chem} activates channel processing only when at least one receiver is associated with that frequency; otherwise, samples on that channel are not processed.

	      The \gls*{uhd} \gls*{api} provides two transmission modes: {\em scheduled} and {\em unscheduled}.
	      In unscheduled mode, TX metadata does not carry a requested transmit time; therefore, each frame is appended to an on-the-air \gls*{fifo} and assigned the next available transmission instant.
	      Following the first unscheduled call, subsequent calls are serialized at the end of the previous transmission.
	      In scheduled mode, TX metadata includes an explicit hardware transmit time, and frames are inserted into the \gls*{fifo} accordingly.

	      To emulate \gls*{usrp} timing semantics, \gls*{vusrp} evaluates each transmit request against a virtual hardware clock derived from the emulation computer clock.
	      For scheduled transmissions, the gap between the current virtual hardware time and the requested transmit time is evaluated at each transmit request.
	      If a scheduled frame specifies a transmit time earlier than the current virtual hardware time, the request is marked {\em late} and the frame is discarded.
	      An \textit{underflow}/\textit{underrun} condition is treated separately and corresponds to the transmitter not being able to sustain continuous sample delivery when samples are needed on the emulated radio timeline.
	      Conversely, if the application produces samples excessively ahead of consumption, such that this gap exceeds the user-defined buffer threshold (in milliseconds), subsequent \gls*{uhd} calls are blocked to emulate hardware backpressure.

	\item[Receiving:]
	      A \gls*{vusrp} receiver is instantiated when a \gls*{usrp} application requests an RX stream through the \gls*{uhd} \gls*{api}.
	      The receiver first requests a data-path port from \gls*{chem}, and then starts a UDP server for receiving signal frames.
	      Receive processing is organized in two stages.

	      In the first stage, a {\em timed buffer} aligns and superposes incoming frames.
	      A receiving \gls*{vusrp} may observe partially overlapping frames from multiple transmitting \glspl*{vusrp}, including itself when transmitter(s) operate on the same frequency.
	      As shown in Fig.~\ref{fig:timed_buffer}, each incoming frame is inserted according to its emulated transmit time: non-overlapping frames are delivered at scheduled receive times as buffer time advances, whereas overlapping frames are combined by sample-wise summation over the overlap interval.
	      The timed buffer is implemented as a priority queue to maintain low latency and overhead during real-time operation.

	      \begin{figure}[H]
		      \centering
		      \includegraphics[width=\columnwidth]{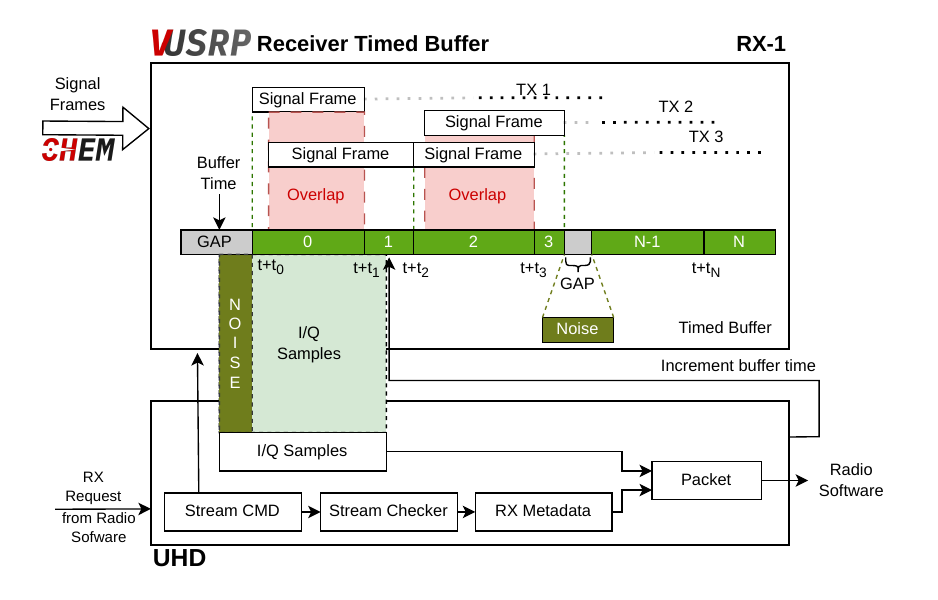}
		      \caption{The timed buffer data structure is used to combine (superpose) signal frames from different nodes arriving at different times and then delivers them to the application software as requested.}
		      \label{fig:timed_buffer}
	      \end{figure}

	      The second stage is triggered when application software requests samples with RX metadata.
	      Consistent with transmission, reception supports unscheduled and scheduled modes.
	      Unscheduled reception provides continuous sample delivery, whereas scheduled reception returns samples at a specified hardware time, a requirement for scheduled communication stacks such as LTE and \gls*{5g}.
	      The receive behavior of a real \gls*{usrp} is replicated using a virtual hardware time, $T_{vusrp}$, which preserves stream continuity and annotates returned samples with reception timestamps.
	      As in real hardware buffering, samples must be consumed to avoid \textit{overflow}; overflow results in lost samples and gaps in the stream.
	      If no transmitter contributes samples within a requested time window, noise-only samples are returned to the application.

	\item[Hardware Time/Clock:] \label{sec:clock}
	      Precise timing is essential for \gls*{lte}, \gls*{5g}, and other similar applications since each frame block has a narrow time window to be transmitted and received.
	      Real \glspl*{usrp} provide the hardware time information through an internal clock, which can be synchronized with external sources such as a \gls*{gpsdo}.
	      In \gls*{vusrp}, the virtual hardware clock is initialized through \gls*{uhd} and derives its time from the clock of emulation host.
	      The assumption is that all \glspl*{vusrp} in a scenario run on the same emulating computer (and are thus precisely synchronized).
	      The virtual clock represents the time with 1~ns resolution.
	      This resolution does not specify synchronization accuracy or bound operating-system scheduling jitter.
	      In the current implementation, this constrains the maximum supported sample rate to 1~GSamples/s.
	      Time resolution limits on real \glspl*{usrp} are imposed by master clock rate.

\end{description}

\subsection{CHEM} \label{sec:achem_imp}

\begin{figure*}[ht]
	\centering
	\includegraphics[width=\textwidth]{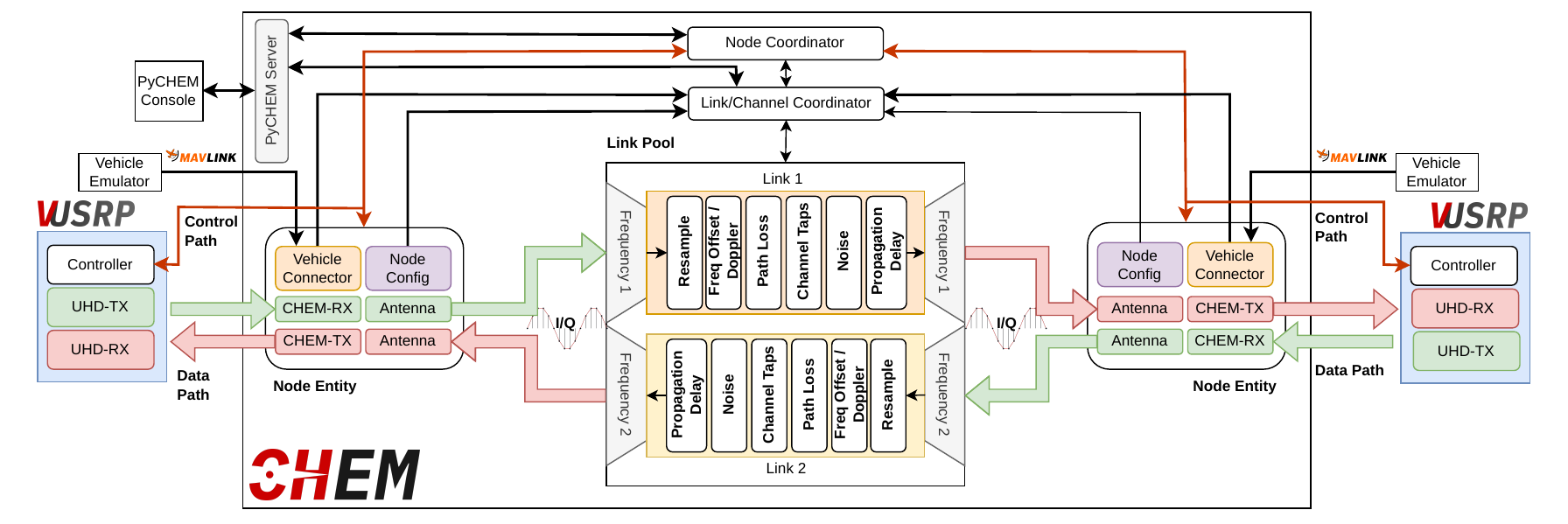}
	\caption{CHEM internal architecture and flow of the wireless channel emulation for two V-USRPs on different uplink and downlink frequencies with a single RX and TX stream each.}
	\label{fig:achem_arch}
\end{figure*}

The main role of CHEM in the emulation system is to replicate wireless channels in real systems.
\gls*{chem} models radio-signal propagation between multiple transmitters and receivers, applies impairments, and resamples signals when needed for the target receiver.
\gls*{chem} coordinates the wireless channels and the \glspl*{vusrp}.
The detailed internal structure of \gls*{chem} is shown in Fig.~\ref{fig:achem_arch}.
\gls*{chem} comprises multiple components that replicate real-world hardware entities in the digital domain while preserving the integrity of system components.
\gls*{chem} is written in C++ and all data path processing is implemented with \gls*{simd} instructions to maintain low latency across the system.

\begin{description}[style=unboxed,leftmargin=0cm]
	\item[Node Entity:]
	      \gls*{chem} assigns a Node Entity to each registered \gls*{vusrp}, including CHEM-TX, CHEM-RX, and a Vehicle Connector (if the node is mobile).
	      The data path of the \gls*{vusrp} is used to exchange I/Q samples between the Node Entity and its corresponding \gls*{vusrp}.
	      The control path is shared between \gls*{vusrp} instances and it is used for coordinating \gls*{vusrp} operation.

	      CHEM-RX is the receiver component of the Node Entity, which uses a UDP server for receiving signal frames from a \gls*{vusrp}.
	      CHEM-RX is instantiated when the ports are assigned for the \gls*{vusrp} and only \emph{one} CHEM-RX is instantiated per \gls*{vusrp}, including for \glspl*{vusrp} with MIMO streams (as the I/Q samples are multiplexed by the \gls*{vusrp} in a single signal frame).
	      The received signal frames are first converted into CPU format and then placed in the attached channel's buffer.

	      CHEM-TX is the transmitter component that uses a UDP client for transferring signal frames from \gls*{chem} to \gls*{vusrp}.
	      The UDP client starts sending samples when the \gls*{chem} assigns a port for the \gls*{vusrp}.
	      Only one CHEM-TX is assigned for each \gls*{vusrp} instance and it holds the related information.
	      CHEM-TX has a \gls*{fifo} buffer that stores frames forwarded by the attached channel.
	      Each item in the buffer is processed according to the receiving \gls*{vusrp}, including receiver antenna gain and \gls*{otw} format change.

	      In addition to CHEM-TX and CHEM-RX, the Node Entity includes the Vehicle Connector.
	      Based on the configuration of the corresponding node, a \gls*{vusrp} can be attached to a fixed or portable node.
	      If the \gls*{vusrp} is configured to be a portable node, the Vehicle Connector initiates a connection with the corresponding MAVLink source, ArduPilot \gls*{sitl} vehicle emulator in our case, and starts receiving vehicle updates for the node entity, including vehicle position and orientation in 3D space.

	\item[Node Coordinator:]
	      The Node Coordinator handles new \glspl*{vusrp} registration, updates of node information, channel requests, and detachment.
	      The Node Coordinator uses a TCP server to receive JSON-based control requests from \glspl*{vusrp}.
	      Node Entities are created by the Node Coordinator and their management is also handled by the Node Coordinator.

	      A \gls*{vusrp} registers itself to the \gls*{chem} with node name and number and type of \gls*{rf} port (RX, TX, or RX/TX).
	      Each \gls*{vusrp} receives two port numbers, which are used for transmitting and receiving data paths.
	      Additionally, each \gls*{vusrp} receives an \gls*{uuid} for further control path communications with the \gls*{chem}, at which point, the \gls*{vusrp} becomes registered with the Node Coordinator.

	\item[Link/Channel Coordinator:]
	      When the application software requests the instantiation of an RX or TX stream, the V-USRP provides the corresponding stream information (e.g., frequency, sampling rate, number of channels) to \gls*{chem}.
	      Upon receipt of this information at \gls*{chem}, a Channel is created by the Link/Channel Coordinator for that particular center frequency if none already exists.
	      If a Channel already exists for that particular frequency, the \gls*{vusrp} stream is assigned to that existing Channel, which leads to creating links between others already attached to that Channel and the new \gls*{vusrp} stream.
	      The Channel starts processing the I/Q samples after at least one receiver and one transmitter are associated with the Channel.
	      A \gls*{vusrp} can get detached from a Channel by either changing the operating center frequency or closing the respective stream.
	      The Channel gets destroyed if all the associated \gls*{vusrp} instances are detached from that particular frequency.

	\item[CHEM Channel and the Link Pool:]
	      The processing of an I/Q sample through \gls*{chem} starts at the CHEM-RX, which receives signal frames from \glspl*{vusrp}.
	      Each CHEM-RX receiver runs on a separate thread and puts the signal frames into the respective Channel pipeline as they are received from the \gls*{vusrp} transmitter.
	      The Node Coordinator also updates the Channel with vehicle information (position and orientation), as well as antenna patterns.
	      Each Channel keeps track of the attached CHEM-RX receivers and CHEM-TX transmitters as they attach, detach, or update their stream parameters.
	      \gls*{chem} uses this information to calculate the parameters of individual channels or links between each receiver and transmitter.

	      In the Channel, I/Q processing starts with the retrieval of CHEM-RX receiver and CHEM-TX transmitter information including mobility if present.
	      For each TX/RX node pair on a Channel, node location, relative distance, \gls*{aoa}, \gls*{aod}, and path loss are calculated.
	      In the Channel, I/Q samples are resampled if required by the CHEM-TX stream options, and transmitter antenna gain is applied according to the \gls*{aod}.
	      Then, the frequency offset, path loss, channel taps, noise, and propagation delay are applied accordingly.
	      Formally, the per-receiver baseband signal model used by \gls*{chem} can be written as:
	      \begin{equation}
		      \label{eq:achem_signal_model}
		      \begin{split}
			      y_r[n] = \sum_{t \in \mathcal{T}_r} G_{t,r}^{\mathrm{rx}}[n] \cdot a_{t,r}[n]
			      \cdot \sum_{\ell=0}^{L_{t,r}-1} h_{t,r,\ell}[n] \cdot G_{t,r}^{\mathrm{tx}}[n] \\
			      \cdot x_t[n-d_{t,r}[n]-\ell] \cdot e^{j\omega_{t,r}n}
			      + w_r[n],
		      \end{split}
	      \end{equation}
	      where $x_t[n]$ and $y_r[n]$ are transmitted and received complex baseband samples, $h_{t,r,\ell}[n]$ is tap $\ell$ of the time-varying \gls*{cir} between transmitter $t$ and receiver $r$, $a_{t,r}[n]$ is the path-loss attenuation, $d_{t,r}[n]$ is the propagation delay in samples, $\omega_{t,r}=2\pi\Delta f_{t,r}T_s$ is the discrete-time frequency offset, $G_{t,r}^{\mathrm{tx}}[n]=g_t^{\mathrm{tx}}(\phi_{t,r}[n])$ and $G_{t,r}^{\mathrm{rx}}[n]=g_r^{\mathrm{rx}}(\theta_{t,r}[n])$ are antenna gains from \gls*{aod}/\gls*{aoa}, and $w_r[n]$ is additive noise.
	      As reflected in \eqref{eq:achem_signal_model}, receiver antenna gains (from \glspl*{aoa}) are applied before each processed I/Q frame is enqueued in the corresponding CHEM-TX buffer.
	      Finally, each processed signal frame is distributed to CHEM-TX transmitters and sent to \gls*{vusrp} receivers (UHD-RX) in parallel.

	      By default, \gls*{chem} provides two measured antenna patterns used in our testbed: the SA-4400-5900 C-band stub antenna (portable node) and the RW-WB1-DN sub-6 surface-mount antenna (fixed node).
	      It also includes standard reference patterns, such as isotropic and dipole models.
	      This information is combined with the vehicle's position information relative to the other node, and the corresponding antenna gain values are extracted.
	      Any new antenna patterns can be added to the system by the user.

	\item[Node Mobility:]
	      One of the important features that \gls*{achem} offers is support for node mobility.
	      Nodes can be attached to a vehicle or placed at a fixed position through PyCHEM or the configuration file associated with a \gls*{vusrp}.
	      A vehicle can be a \gls*{uav} or a \gls*{ugv}, emulated through a MAVLink source (ArduPilot \gls*{sitl} vehicle software in this work).
	      The ArduPilot \gls*{sitl} emulator initiates the MAVLink connection the same way it does for real vehicles.
	      \gls*{chem} uses the \gls*{sitl}'s TCP MAVLink connection to receive status messages from the emulated vehicle.
	      Although ArduPilot is used in this work, \gls*{chem} can operate with any MAVLink-compatible vehicle emulator or tool that sends MAVLink messages.
	      The number of vehicles, their node associations, and fixed-node positions are obtained from the \gls*{chem} configuration file.
	      In addition to providing updates to \gls*{chem}, the vehicle \gls*{sitl} interacts with the application vehicle software (as shown in Fig.~\ref{subfig:achem_system}) for automation and control.

	\item[PyCHEM:]
	      PyCHEM provides the management interface and control \gls*{api} for \gls*{chem} via a TCP connection.
	      Through PyCHEM, an experimenter can configure and update channel- and node-level parameters, including antenna models, frequency offsets, noise and propagation models, fixed-node locations, and portable-node vehicle types.
	      This interface enables online reconfiguration of active experiments without restarting the emulation framework.
\end{description}

\section{Validation}

\begin{figure}[ht]
	\centering
	\includegraphics[width=\columnwidth]{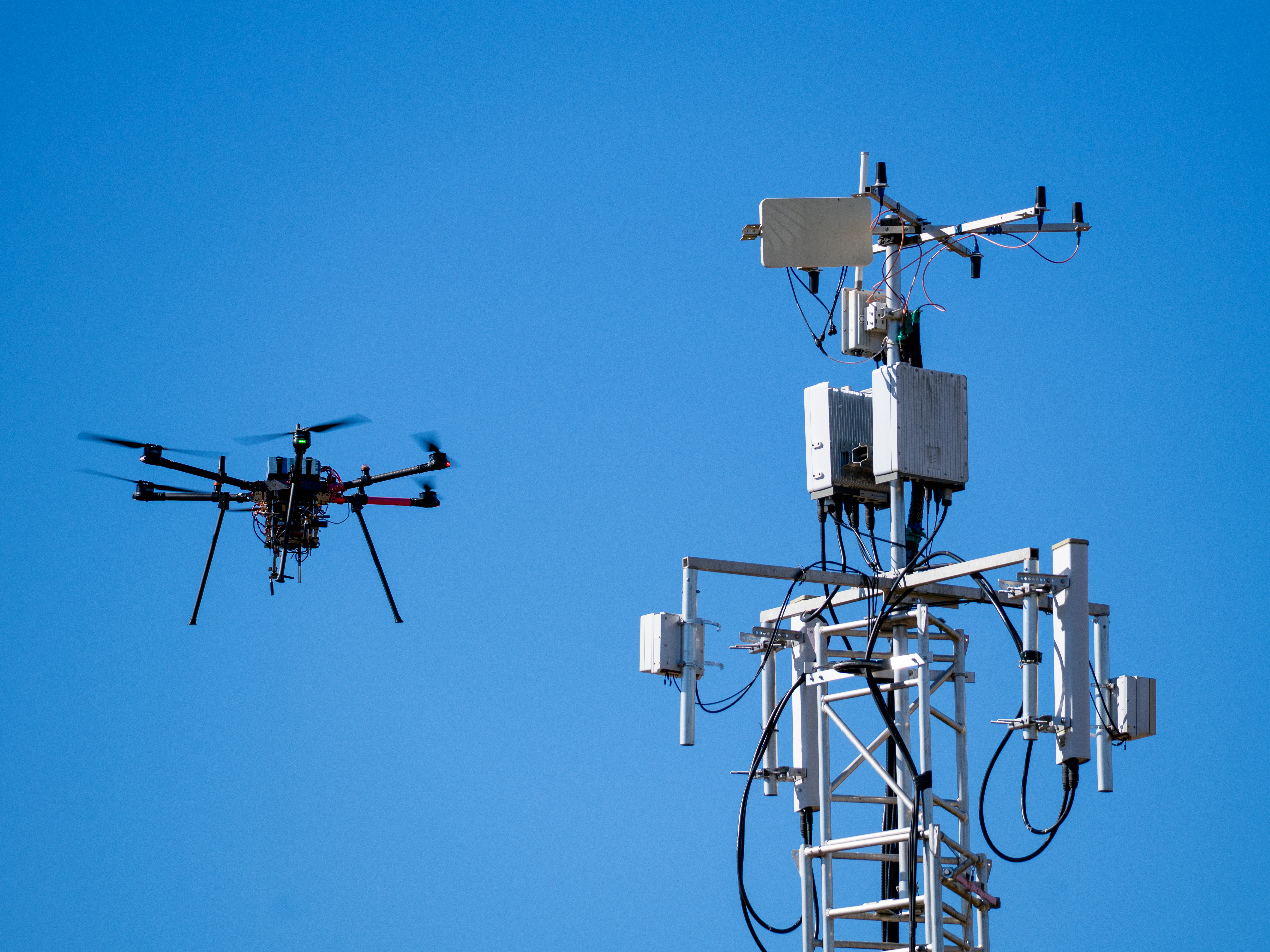}
	\caption{Experimental setup with a fixed node and a Large AERPAW Multicopter (LAM) at AERPAW's Lake Wheeler testbed site.}
	\label{fig:achem_validation}
\end{figure}

We validate \gls*{achem}'s ability to run unmodified cellular software and reproduce system-level behavior observed in over-the-air experiments at AERPAW's Lake Wheeler testbed.
Using matched trajectories and stack configurations, we compare LTE link quality and throughput, mobility driven LTE handover, and 5G \gls*{nr} results obtained with OpenAirInterface stack.
These comparisons assess fidelity under the selected channel model and experimental configurations.
The development and validation of higher fidelity propagation models are outside the scope of this work.

\subsection{Experimental Setup}

\begin{figure}[ht]
	\centering
	\includegraphics[width=\columnwidth]{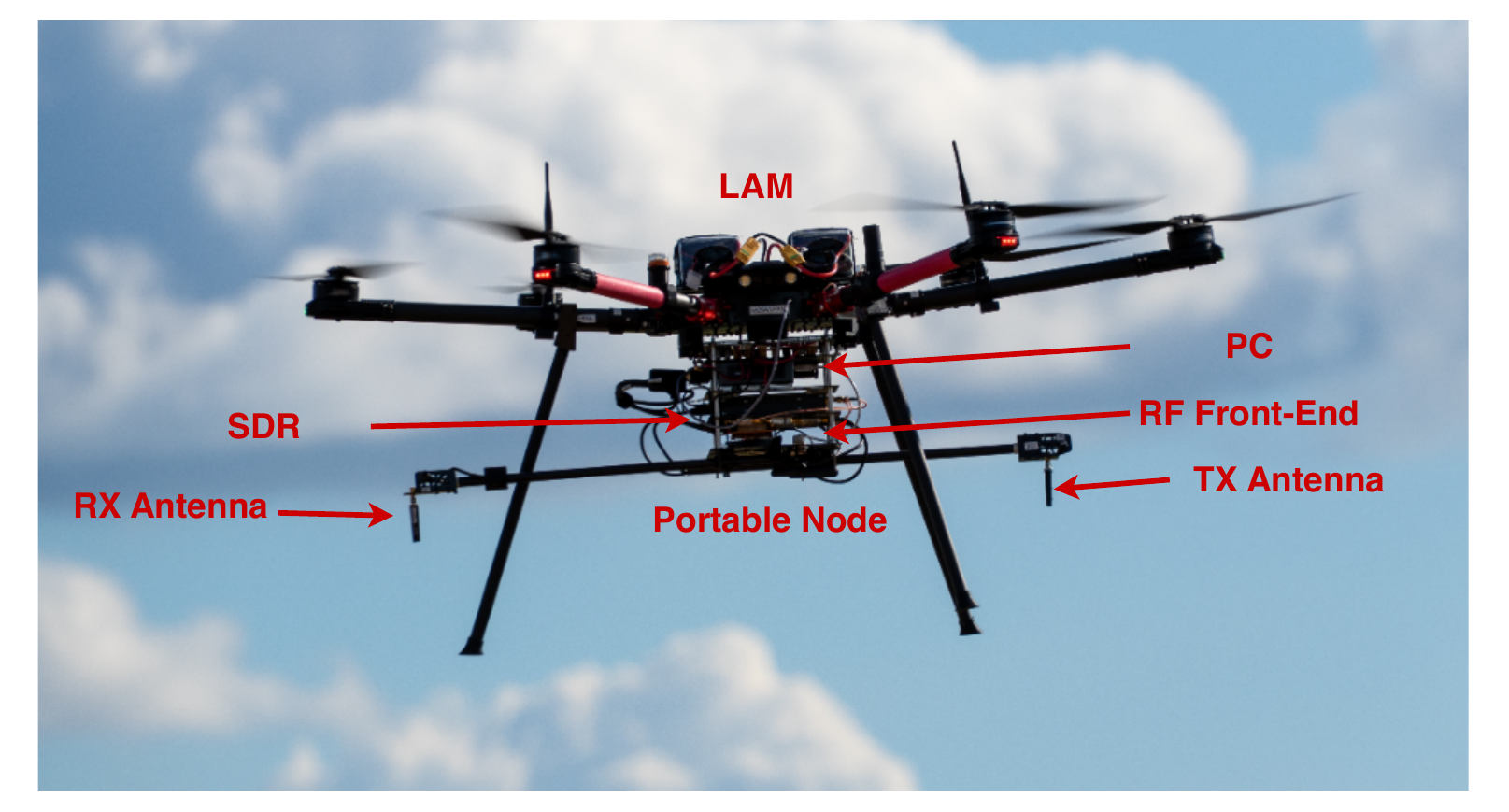}
	\caption{LAM carrying the portable node used in the experiments.}
	\label{fig:lam_pn}
\end{figure}

The experiments were conducted at AERPAW's Lake Wheeler testbed and in its \gls*{achem}-based digital twin.
The physical setup consists of a tower-mounted fixed node equipped with a server and a \gls*{usrp} B210, and a portable node carried by a Large AERPAW Multicopter (LAM), shown in Figs.~\ref{fig:achem_validation} and~\ref{fig:lam_pn}.
The fixed node antenna is mounted on a tower with a height of 12~m.
The portable node comprises an Intel NUC, a \gls*{usrp} B210, shown in Fig.~\ref{fig:achem_validation}, and an \gls*{rf} front end with transmit and receive amplifiers and a 3000--4300~MHz bandpass filter.

In \gls*{achem}, the nodes and their wireless links are represented by \glspl*{vusrp} and \gls*{chem}, respectively.
The emulation uses antenna radiation patterns corresponding to the fixed and portable node antennas, with \gls*{uav} mobility emulated by ArduPilot \gls*{sitl}~\cite{ardupilot}.
The physical and emulated experiments use the same application software (srsRAN and OpenAirInterface) and configurations.
In both environments, each node's application software runs without modification in an Ubuntu 22.04-based Docker container, facilitating deployment between the testbed and digital twin illustrated in Fig.~\ref{fig:system}.

During the experimentation in digital twin, wireless propagation is represented by a two-ray ground reflection model comprising a direct patch and a single ground reflected path.
This model was selected to approximate propagation between the tower mounted and \gls*{uav} mounted antennas in the open flight area, assuming a flat reflecting ground surface.
Path loss is computed from the coherent sum of the two contributions using their geometric path lengths and carrier frequency.
The tower height is set to same height as the real counterpart, while the \gls*{uav} height and horizontal distance are derived from the vehicle position supplied by \gls*{sitl} relative to the fixed node.
A constant ground reflection coefficient of $\Gamma=-0.3$ is used, selected using prior channel measurements that were independent of the cellular validation data.

Each case study was repeated over multiple runs in the physical testbed, and measurements from these runs were used to quantify performance.
For clarity, the time-series figures present the testbed results for each case study alongside the corresponding \gls*{achem} results, highlighting the stack operation and system level behavior across two environments.

\begin{figure*}[!ht]
	\centering
	\subfloat[\gls*{lte} \gls*{snr} comparison between the testbed and \gls*{achem}.\label{fig:4g_snr}]{%
		\includegraphics[width=\columnwidth]{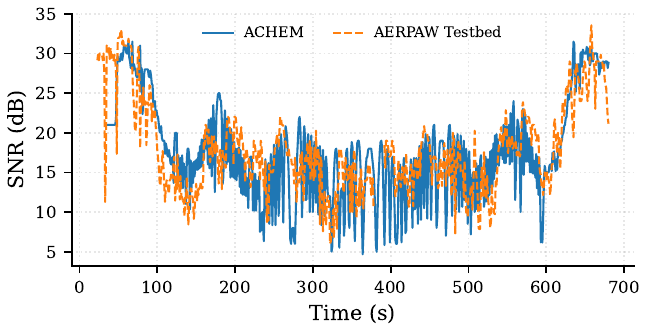}
	}
	\hfill
	\subfloat[LTE RSRP comparison between the testbed and \gls*{achem}.\label{fig:4g_rsrp}]{%
		\includegraphics[width=\columnwidth]{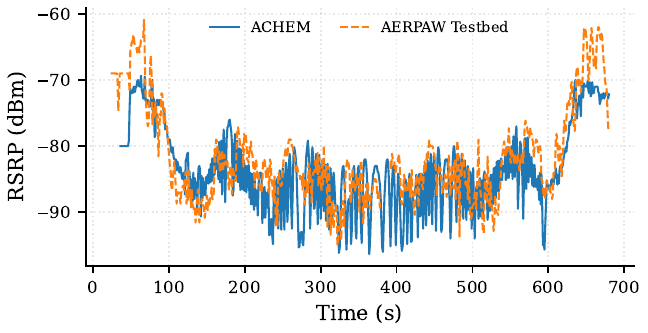}
	}
	\caption{LTE link-quality traces collected during the same mobility flight.}
	\label{fig:lte_snr_rsrp}
\end{figure*}

\subsection{Case Study: LTE} \label{subsec:lte}

\begin{figure}[h]
	\centering
	\includegraphics[width=\columnwidth]{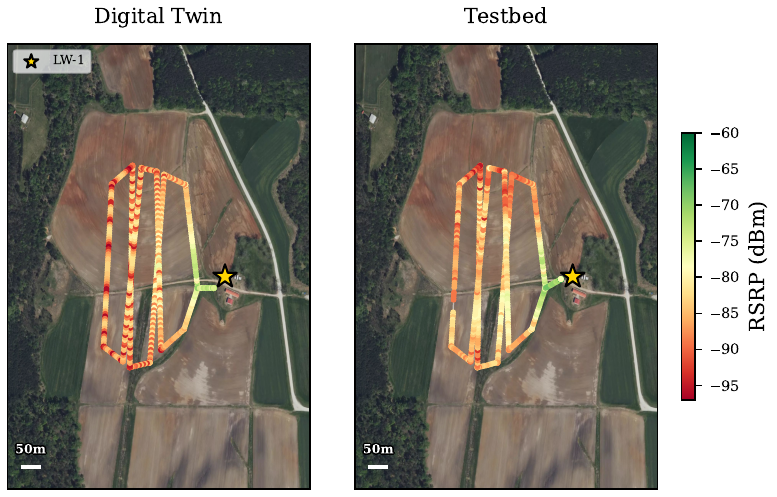}
	\caption{\gls*{uav} zigzag trajectory and corresponding \gls*{rsrp} values in dBm from the \gls*{lte} validation experiment.}
	\label{fig:lte_zigzag_trajectory}
\end{figure}

We evaluate end-to-end operation of the srsRAN \gls*{lte}~\cite{srslte,srslte_2} stack in \gls*{achem} and the physical testbed with a mobile \gls*{ue} deployed on a \gls*{uav}.
The experiment exercises the integration of the cellular applications, \gls*{vusrp} radio interfaces, and \gls*{chem}'s mobility dependent channel processing.
srsRAN is widely used in research testbeds \cite{sdr_testbed,sdr_testbed_oran,powder_tb} and provides end-to-end \gls*{lte} functionality through srsUE, srsENB, and srsEPC.

For this experiment, the \gls*{lte} downlink center frequency is set to 3410~MHz and the uplink center frequency is 3320~MHz.
The \gls*{enb} and \gls*{epc} run on the fixed node, while the \gls*{ue} runs on the portable node.
The \gls*{lte} bandwidth is 10~MHz (50 \glspl*{prb}) and the sampling rate is 11.52~MHz, consistent with \gls*{lte} specifications \cite{etsi_ts_136_101}.

\begin{figure}[h]
	\centering
	\includegraphics[width=0.96\columnwidth,trim={0 0 0 0.1cm},clip]{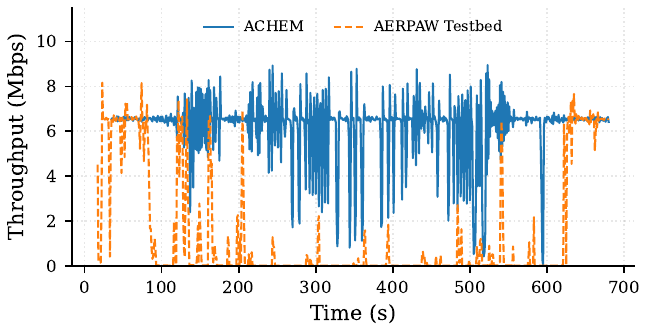}
	\caption{\gls*{lte} uplink throughput comparison between the testbed and \gls*{achem}.}
	\label{fig:lte_throughput_comparison}
\end{figure}

\begin{figure}[h]
	\centering
	\includegraphics[width=0.96\columnwidth,trim={0.1cm 0 0 0.1cm},clip]{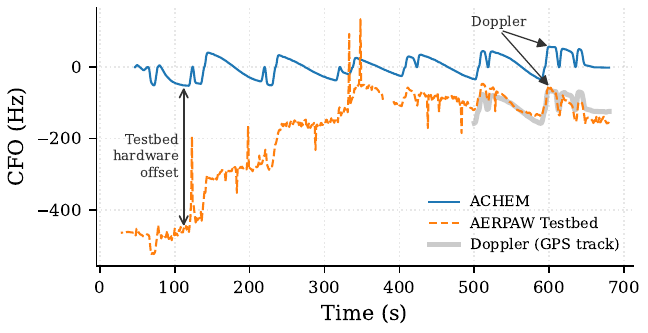}
	\caption{\gls*{lte} \gls*{cfo} comparison between the testbed and \gls*{achem}.}
	\label{fig:lte_cfo_comparison}
\end{figure}

\begin{figure}[ht]
	\centering
	\includegraphics[width=0.85\columnwidth]{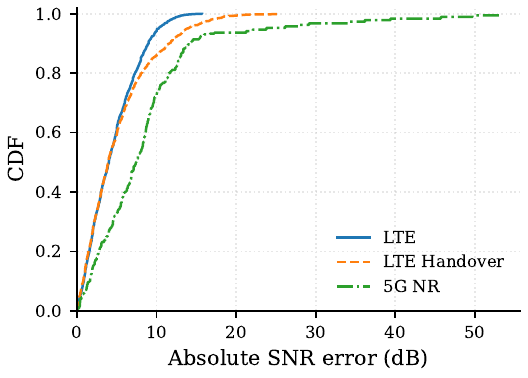}
	\caption{Empirical \glspl*{cdf} of absolute SNR differences between ACHEM and the AERPAW testbed for LTE, LTE handover, and 5G NR.}
	\label{fig:snr_error_cdf}
\end{figure}

The flight lasts approximately 10~minutes along a rectangular trajectory at 30~m altitude and 5~m/s horizontal speed.
Figs.~\ref{fig:4g_snr} and~\ref{fig:4g_rsrp} compare the reported SNR and RSRP during the flight along the \gls*{uav} trajectory shown in Fig.~\ref{fig:lte_zigzag_trajectory}.
TCP uplink traffic from \gls*{ue} (client) to \gls*{enb} (server) is generated with MGEN \cite{mgen}, with observed \gls*{ul} throughput in the 10~Mbps range in both \gls*{achem} and the testbed.
No {\em overflows} or {\em underflows} were observed in either run.
Fig.~\ref{fig:lte_throughput_comparison} shows the corresponding \gls*{lte} uplink throughput comparison between the over-the-air testbed and \gls*{achem}.
Both environments exhibit similar overall variations during the flight, with higher link quality near the beginning and end, where \gls*{uav} is closer to the eNB.
The empirical CDF in Fig.~\ref{fig:snr_error_cdf} quantifies the absolute SNR difference between the testbed and \gls*{achem}.
The median absolute difference is approximately 4~dB, and approximately 90\% of the evaluated differences are below 9~dB.

Fig.~\ref{fig:lte_cfo_comparison} compares the reported \gls*{cfo} and includes a Doppler reference derived from the \gls*{uav}'s GPS trajectory.
The \gls*{achem} trace exhibits time-varying offsets around zero, whereas the testbed trace contains a larger offset.
During the latter portion of the flight in testbed run, \gls*{usrp}'s hardware clock warmed up and the doppler effect became more apparent as highlighted in Fig.~\ref{fig:lte_cfo_comparison}

\subsection{Case Study: LTE Handover}
\begin{figure}[h]
	\centering
	\includegraphics[width=\columnwidth]{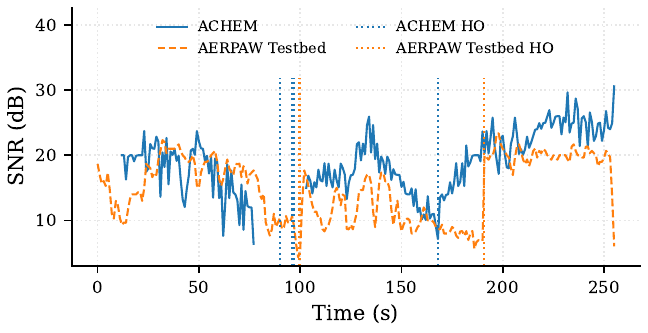}
	\caption{SNR comparison and handover (HO) events during LTE HO experiments in the testbed and \gls*{achem}.}
	\label{fig:handover_snr}
\end{figure}

We extend the \gls*{lte} experiment with a second eNB at Lake Wheeler Tower~2 to demonstrate handover process between cells.
Since srsRAN's \gls*{epc} core network (srsEPC) lacks S1 handover support, we replace srsEPC with Open5GS~\cite{open5gs} as core network for this experiment.
The eNBs at Towers~1 and~2 use \glspl*{pci} 1 and 6, respectively.
The channel bandwidth configuration remains 10~MHz~(50~\glspl{prb}).
Both environments use the same handover configuration; A3 offset is set to 6, trigger time is 480~ms, and the corresponding root sequence indices are 204 and 264 for the two cells.

Handovers complete successfully in both environments, and uplink ping traffic continues during the flight.
Fig.~\ref{fig:handover_snr} shows the reported SNR and recorded handover events.
Both traces shows declining link quality before the handover and recovery afterwards.
Residual \gls*{snr}-level gaps are expected from unmodeled platform-specific effects, such as RF front-end characteristics and cable losses.
The LTE handover curve in Fig.~\ref{fig:snr_error_cdf} summarizes the absolute differences between the testbed and \gls*{achem} SNR measurements.
Compared to the single cell \gls*{lte} experiment, distribution shows larger SNR difference due to the timing of the handover.

\subsection{Case Study: 5G}
\begin{figure}[h]
	\centering
	\includegraphics[width=\columnwidth]{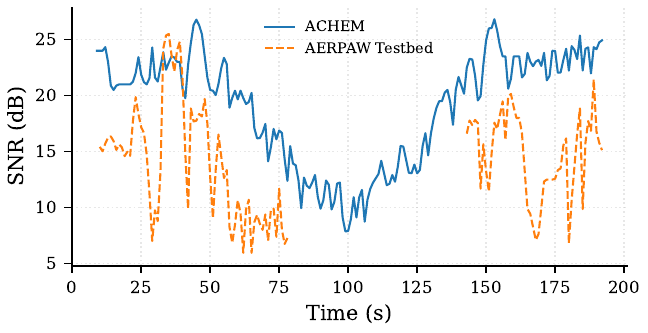}
	\caption{OpenAirInterface~\cite{oai_5g} \gls*{5g} \gls*{snr} comparison between the testbed and \gls*{achem}.}
	\label{fig:5g_snr}
\end{figure}

To further validate \gls*{achem} with a more complex stack, we extend the evaluation to \gls*{5g} NR using OpenAirInterface (OAI)~\cite{oai_5g} \gls*{gnb} and \gls*{ue} stacks.
This experiment demonstrates operation with a different cellular stack implementation and TDD configuration.
The center frequency of the gNB is 3410~MHz, the channel bandwidth is 20~MHz, and the configured sampling rate is 23.04~MHz.
The TCP \gls*{ul} traffic from \gls*{ue} to \gls*{gnb} is generated through MGEN \cite{mgen}.
The \gls*{5g} experiment is performed with the same hardware configuration as the \gls*{lte} experiment but with a shorter trajectory (approximately 2~minutes) due to link reliability challenges with the OAI \gls*{5g} stack.

Fig.~\ref{fig:5g_snr} compares the reported SNR in the testbed and \gls*{achem}.
Over the experiments, \gls*{achem} generally shows higher SNR, while the testbed measurements show more pronounced short-term fluctuations.
In the testbed run, the UE loses its connection and then reconnects resulting in a gap in the reported SNR.
The 5G NR curve in Fig.~\ref{fig:snr_error_cdf} summarizes the absolute SNR difference.
An uplink throughput around 10~Mbps is observed in both \gls*{achem} and the testbed.
This can be further improved by tuning the channel model parameters and incorporating additional impairments, such as phase noise and non-linearities, which are not currently modeled in \gls*{achem}.

\subsection{Limitations}

The agreement between an emulated scenario and its physical counterpart depends on the selected propagation model, its parameters, and the radio effects represented.
The two-ray model used in these experiments approximates the evaluated environment, but does not capture all site-specific propagation effects.
\gls*{achem} also does not explicitly model power-amplifier nonlinearities and compression, receiver \gls*{agc} dynamics and saturation, oscillator phase noise and long-term drift, or detailed analog front-end effects such as I/Q imbalance and filter responses.
These omissions, together with propagation-model approximations, contribute to the observed differences in link quality, frequency offset, and throughput.
\gls*{achem} has only been tested with all \glspl*{vusrp} running on the same emulation host to enable time synchronization between \glspl*{vusrp}.

\section{Performance Evaluation}
To support real-time operation, ACHEM must process I/Q samples within application timing constraints while keeping end-to-end latency low.
Since ACHEM consists of two main runtime components, \gls*{vusrp} and \gls*{chem}, we evaluate them separately.
We use the \gls*{chem} and \gls*{vusrp} benchmark utilities to measure end-to-end latency, per-stage \gls*{dsp} latency, throughput, and drop rate using synthetic I/Q signals.
The \gls*{vusrp} benchmark characterizes the endpoint radio reception emulation overhead, specifically superposition, while the \gls*{chem} benchmark evaluates the scalability and DSP cost of the channel processing pipeline.
The performance evaluation combines link count scalability sweeps (Fig.~\ref{fig:multi_node_performance}) using the LTE/NR sample rates and frame durations.

\subsection{Benchmark Setup}
\gls*{achem}'s limits are evaluated on two differrent hosts; a server equipped with an AMD EPYC 9534 (64 cores) and 512~GB of RAM, and a laptop with Intel Core i7 12700H (14 cores) and 64~GB of RAM running.
The server runs a real-time Linux kernel, whereas the laptop runs a standard Linux kernel.
These configurations represents two edges in the spectrum, which shows \gls*{achem}'s performance and limitations on different computers.
During the benchmark, each host had minimal system load and processor affinity is applied to the \gls*{chem} threads.

\subsubsection{V-USRP Scalability}
\begin{figure}[h]
	\centering
	\includegraphics[width=\columnwidth]{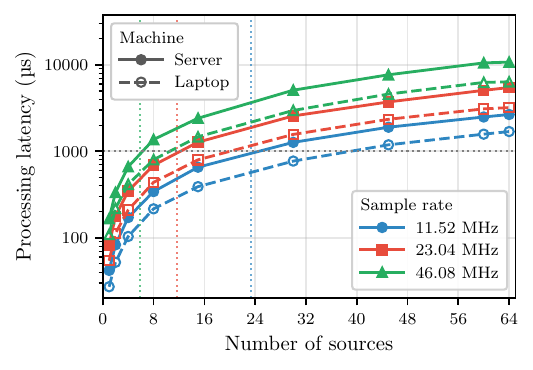}
	\caption{V-USRP end-to-end latency as a function of the number of concurrent sources (i.e., transmitters) for a single receiver.}
	\label{fig:vusrp_latency}
\end{figure}

The \gls*{vusrp} benchmark measures receiver-side processing latency independently of \gls*{chem}'s channel processing workload.
A single \gls*{vusrp} receiver is instantiated and configured to receive from $N$ concurrent sources, where $N$ varies from 1 to 64.
Each source injects pre-generated I/Q frames into the receiver's timed buffer at scheduled times, using the LTE/NR sampling rates.
This benchmark measures the cost of accepting, ordering, and superposing incoming frames as the number of contributing sources increases.
Fig.~\ref{fig:vusrp_latency} shows the measured \gls*{vusrp} latency as a function of the number of concurrent sources for a single receiver.
The results show increasing processing latency with both source count and sampling rate.
At a fixed sampling rate, additional sources increase the number of overlapping frmaes that the receiver must combine/superpose.
At 11.52MHz sampling rate, latency remains below 1~ms for up to 24 concurrent sources for both hosts, which is sufficient for many LTE/NR scenarios, and it remains below 2~ms for up to 64 concurrent sources.
Considering the scheduling limititations on LTE/NR stacks, where frames are scheduled 4 ms ahead (varies on 5G NR), \gls*{vusrp} supports up to 10 nodes for LTE/NR scenarios under given host configurations.
This is later confirmed by running experiment with srsRAN LTE stack individually with one eNB and eight UEs.

The host comparison shows that receive-processing latency depends on the tested platform configuration.
During the comparison between laptop and server, the laptop exhibited lower latency than the server across the configurations used in the benchmark.
The benchmark does not isolate the contributions of processor frequency, memory behavior, and operating-system scheduling.

\subsubsection{CHEM Scalability}
\begin{figure*}[ht]
	\centering
	\subfloat[CHEM Processing latency versus number of nodes (N).\label{fig:e2e_latency}]{%
		\includegraphics[width=0.49\textwidth,trim={0 0 0.1cm 0.1cm},clip]{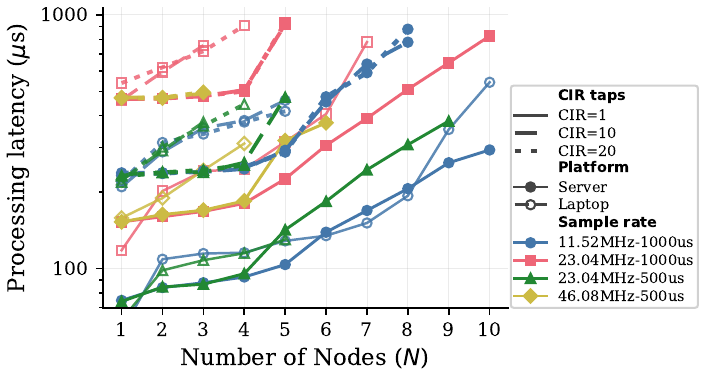}
	}
	\hfill
	\subfloat[CHEM Signal Frame drop rate versus number of emulated links.\label{fig:drop_rate}]{%
		\includegraphics[width=0.49\textwidth,trim={0 0 0.1cm 0.1cm},clip]{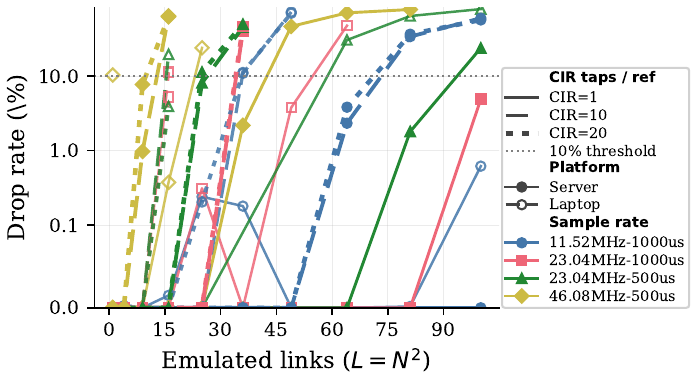}
	}
	\caption{CHEM performance on the server and laptop across the evaluated sample-rate/frame-duration profiles and CIR tap counts.}
	\label{fig:multi_node_performance}
\end{figure*}
\gls*{chem} benchmark evaluates \gls*{chem}'s scalability by measuring latency, throughput, and drop rate under increasing load configurations.
For each scenario, the benchmark instantiates $N$ transmitters and $N$ receivers~($N^2$ links) connected through \gls*{chem}'s channel processing pipeline.
Synthetic I/Q signal frames are injected into mocked \gls*{chem}-RX queues sequentially.
The injection interval determines the offered frame rate, while sampling rate and frame duration determine the number of samples per frame.
Depending on the configuration, processing pipeline includes \gls*{cir} convolution, path-loss attenuation, noise generation and injection, antenna effects, and frequency-offset/Doppler processing before the frames reach \gls*{chem}-TX.
The benchmark reports end-to-end latency (with per-stage breakdown) and drop rate (i.e., signals whose intended transmission time passes the maximum acceptable threshold, by default 5~ms, configurable based on application requirements).
The benchmark records latency within the instrumented \gls*{chem} pipeline, per-stage processing times, and deadline-based frame drops.
A run is considered successful when the drop rate remains below a configurable threshold (10\%, which is a common working threshold for LTE/5G systems from our tests).
Each measurement phase is preceded by a warm-up of 1{,}000 signal frames (each is 1~ms long) to stabilize the caches and thread pools.
\begin{table}[t]
	\vspace{-1.2\baselineskip}
	\centering
	\caption{Latency statistics for 3GPP-realistic single-link configurations ($L=1$). Latency is reported in $\mu$s as mean values.}
	\label{tab:latency_breakdown}
	\setlength{\tabcolsep}{3pt}
	\footnotesize
	\resizebox{\columnwidth}{!}{%
		\begin{tabular}{lrrrr}
			\toprule
			                                    & \multicolumn{1}{c}{11.52MHz/1ms} & \multicolumn{1}{c}{23.04MHz/1ms} & \multicolumn{1}{c}{23.04MHz/0.5ms} & \multicolumn{1}{c}{46.08MHz/0.5ms} \\
			\midrule
			\textbf{End-to-end}                 & 82.09                            & 161.44                           & 84.66                              & 167.21                             \\
			\quad Queue wait                    & 7.50                             & 8.74                             & 10.82                              & 14.41                              \\
			\quad Processing                    & 77.28                            & 155.55                           & 76.61                              & 155.62                             \\
			\quad Delivery                      & 0.04                             & 0.05                             & 0.05                               & 0.05                               \\
			\midrule
			\quad CIR                           & 2.77                             & 5.49                             & 2.77                               & 5.49                               \\
			\quad Path loss                     & 0.10                             & 0.17                             & 0.09                               & 0.14                               \\
			\quad Noise Generation \& Injection & 58.49                            & 123.10                           & 58.45                              & 124.33                             \\
			\quad Freq. offset                  & 0.03                             & 0.03                             & 0.03                               & 0.03                               \\
			\midrule
			Throughput (signal frames/s)        & 948                              & 953                              & 1896                               & 1894                               \\
			Drop rate (\%)                      & 0.0                              & 0.0                              & 0.0                                & 0.0                                \\
			\bottomrule
		\end{tabular}
	}
\end{table}

Table~\ref{tab:latency_breakdown} shows single-link ($L=1$) results for four LTE/NR specification-based configurations on the server: LTE-10MHz (11.52~MHz sampling rate, 1~ms frame period), LTE-20MHz (23.04~MHz sampling rate, 1~ms frame period), NR-20MHz (23.04~MHz sampling rate, 0.5~ms slot duration), and NR-40MHz (46.08~MHz sampling rate, 0.5~ms slot duration).
The table reports mean pipeline latency, queuing and per-stage processing durations, and runtimes for individual impairment operations.
Because the runtime is pipelined, the per-stage means should be interpreted as component-level costs and do not sum exactly to the measured end-to-end mean.

Across the four single-link LTE/NR configurations, mean end-to-end latency ranges from 82.09~$\mu$s (11.52~MHz) to 167.21~$\mu$s (46.08~MHz).
Drop rate remains at 0 in all cases, while throughput ranges from 948 to 953~signal frames/s for 1~ms LTE profiles and from 1894 to 1896~signal frames/s for 0.5~ms NR profiles, consistent with the shorter NR injection interval.
Delivery overhead is negligible relative to Queue wait and Processing, and Noise generation/injection is the dominant processing component, as it involves accessing and shuffled access to pre-generated noise samples and applying them to the signal frame.
In our implementation, we chose pre-generated noise over on-the-fly generation to reduce per-frame processing cost.
Fig.~\ref{fig:multi_node_performance} extends this analysis to link count scalability on a server and laptop setups, showing that both processing latency and drop rate increase with link count, with steeper degradation under higher load configurations.
The 10\% reference in Fig.~\ref{fig:drop_rate} marks the practical pass/fail boundary.

\subsection{Practical Operating Limits}

For a fixed host, sample rate, frame duration, channel configuration, and network topology, the maximum supported node count is determined by the end-to-end emulation latency.
For an $N$-node scenario, this latency is described as,
\begin{equation}
	\begin{aligned}
		D_{\mathrm{total}}(N)
		={} & D_{\mathrm{V,TX}}(N) + D_{\mathrm{CHEM}}(N)     \\
		    & + D_{\mathrm{V,RX}}(N) + D_{\mathrm{other}}(N),
	\end{aligned}
\end{equation}
where $D_{\mathrm{V,TX}}$ and $D_{\mathrm{V,RX}}$ denote the transmitting and receiving V-USRP delays, respectively, and $D_{\mathrm{CHEM}}$ denotes the channel-emulation delay under the corresponding network load.
The term $D_{\mathrm{other}}$ accounts for transport, queuing, and scheduling delays in the kernel not already included in the component measurements.
Under the selected real-time budget $T_{\mathrm{RT}}=1\,\mathrm{ms}$, the supported node limit becomes,
\begin{equation}
	N_{\max}
	= \max\left\{
	N\in\mathbb{N}: D_{\mathrm{total}}(N)\leq T_{\mathrm{RT}}
	\right\}.
\end{equation}
Each delay must be evaluated under the same configuration and concurrent workload.
In practice, on the laptop configuration described above, \gls*{achem} successfully supported an srsRAN LTE network with one eNB and eight UEs (nine nodes in total), with no operational issues observed while the UEs actively pushed traffic through the established link.

\section{Related Work}

\begin{table*}[t]
	\centering
	\caption{Comparison of platforms/tools and the \gls*{achem} framework for wireless system development \& testing.}
	\label{tab:sys_comparison}
	\setlength{\tabcolsep}{5pt}
	\renewcommand{\arraystretch}{1.12}
	\scriptsize
	\begin{tabularx}{\textwidth}{@{}>{\raggedright\arraybackslash}X>{\raggedright\arraybackslash}p{4.5cm}cccccc@{}}
		\toprule
		\textbf{Platform} & \textbf{Scope}                  & \makecell{\textbf{Interoper-}                                                     \\\textbf{ability}} & \makecell{\textbf{Hardware}\\\textbf{Emulation}} & \makecell{\textbf{SDR}\\\textbf{Support}} & \makecell{\textbf{Real-}\\\textbf{Time}} & \makecell{\textbf{End-to-End}\\\textbf{Testing}} & \makecell{\textbf{Dynamic}\\\textbf{Channel}\\\textbf{Control}} \\
		\midrule
		NS-3~\cite{ns3}, OMNeT++~\cite{omnet}
		                  & Packet-level network simulation
		                  & $-$                             & $-$                           & $-$        & $-$        & \dag       & \checkmark \\
		Sionna (NVIDIA)~\cite{sionna}
		                  & PHY/link/system simulation
		                  & $-$                             & $-$                           & $-$        & \checkmark & \dag       & \checkmark \\
		Aerial \gls*{dt} (NVIDIA)~\cite{nvidia_aodt}
		                  & NVIDIA 5G/6G digital twin RAN
		                  & $-$                             & $-$                           & $-$        & \dag       & \checkmark & \checkmark \\
		RF-SITL~\cite{rf-sitl}
		                  & GNU Radio
		                  & $-$                             & \dag                          & $-$        & \dag       & \checkmark & \dag       \\
		OAI rfSimulator~\cite{oai_5g}
		                  & OAI NR
		                  & $-$                             & \dag                          & \checkmark & \dag       & \checkmark & \dag       \\
		srsRAN ZMQ~\cite{srslte}
		                  & srsRAN LTE/NR
		                  & $-$                             & \dag                          & \checkmark & \dag       & \checkmark & \dag       \\
		OWDT~\cite{owdtSionna}
		                  & OAI $+$ Sionna 5G mobility
		                  & $-$                             & \dag                          & \checkmark & \checkmark & \checkmark & \checkmark \\
		Hardware-in-the-loop (HITL) Emulator~\cite{peter_emu}
		                  & Any RF Stack
		                  & \checkmark                      & $-$                           & \checkmark & \dag       & \checkmark & \checkmark \\
		PROPSIM (Keysight)~\cite{propsim}
		                  & Any RF Stack
		                  & \checkmark                      & $-$                           & \checkmark & \dag       & \checkmark & \checkmark \\
		Vertex (Spirent)~\cite{vertex_spirent}
		                  & Any RF Stack
		                  & \checkmark                      & $-$                           & \checkmark & \checkmark & \checkmark & \checkmark \\
		\midrule
		\textbf{\gls*{achem}}
		                  & Any UHD-based software
		                  & \checkmark                      & \checkmark                    & \checkmark & \checkmark & \checkmark & \checkmark \\
		\bottomrule
		\vspace{0.5mm}
	\end{tabularx}
	{\footnotesize
	{\checkmark~Fully supported; $-$~Not supported; $\dag$~Partially supported or supported\\[-0.5pt]
		with limitations (e.g., non-real-time execution or support limited to a specific SDR stack).\par
		PHY: physical layer; RAN: radio access network; OWDT: Open Wireless Digital Twin.}
	}
\end{table*}

Various platforms have been developed to support wireless system development and testing, each offering different levels of interoperability, hardware emulation capability, SDR stack support, real-time operation, end-to-end testing support, and dynamic channel control.
Table \ref{tab:sys_comparison} provides a comparative overview of the most relevant solutions, including OAI's rfSimulator, srsRAN ZMQ \cite{srslte}, Sionna \cite{sionna}, Aerial DT \cite{nvidia_aodt}, NS-3 \cite{ns3}, RF-SITL \cite{rf-sitl}, and OWDT \cite{owdtSionna}, alongside the proposed ACHEM framework, while also clarifying the practical scope of each platform.
In this section, we analyze how \gls*{achem} differentiates itself from existing tools by providing software emulation, combining SDR software portability, real-time operation, and enabling full end-to-end testing.

\begin{description}[style=unboxed,leftmargin=0cm]
	\item[Network Simulators:] Network simulators are commonly used for testing the higher layers of communication stacks.
	      ns-3 \cite{ns3} is one of the most well-known network simulators available today that provides simulation capabilities for different communication protocols and techniques (5G mmWave \cite{ns3_mmwave}, Wi-Fi 802.11ax \cite{ns3_80211ax}, etc.).
	      OMNeT++ \cite{omnet} is another network simulator that can model wireless channels with ray-tracing and integration of real-world terrain.
	      A main drawback of these tools is that much of the software in the stack is based on {\em models} of the real networking stacks instead of production code, thus inevitably introducing inaccuracies.
	      Furthermore, for the sake of boosting performance, the physical layer models tend to be simplified, for example acting at the packet level rather than symbol or sample (I/Q) level of accuracy.

	\item[Software-in-the-Loop Channel Emulators:] RF-SITL \cite{rf-sitl} is a software-based channel emulation tool built on top of GNU Radio \cite{gnuradio}.
	      RF-SITL uses University at Buffalo’s Airborne Networking and Communications (UB-ANC) emulator \cite{ub_anc} for emulating mobility.
	      While there are similarities between \gls*{achem} and RF-SITL (e.g., in supporting mobility), RF-SITL only supports GNU Radio-based application software, while \gls*{achem} supports any \gls*{usrp}-based software (including \gls*{5g} cellular networks based on UHD).
	      Furthermore, RF-SITL uses a pre-generated GNU Radio graph connecting nodes together, thus preventing nodes from dynamically joining and leaving during an experiment.

	      With the involvement of AI in the telecommunications industry, NVIDIA developed wireless system testing tools such as Sionna \cite{sionna} and Aerial \cite{nvidia_aodt}.
	      Sionna is a Python library that uses ray-tracing to simulate wireless channels within the specified 3D environment.
	      Since Sionna is simulation-based, the pipeline is abstracted and production level code cannot be used with this tool.
	      There is a study~\cite{owdtSionna} that leverages Sionna's ray-tracing capabilities to generate channel taps and applies them to OAI's internal simulation tool named rfSimulator.
	      However, rfSimulator only works with OAI software and it does not scale with multiple frequencies and other \gls*{sdr}-based applications.
	      On the commercial side, NVIDIA has a digital twin environment called Aerial DT~\cite{nvidia_aodt}, which uses ray-tracing and its own proprietary 5G and 6G stack.
	      Although this is close to what ACHEM offers, it lacks interoperability with other tools and hardware because of the proprietary stack and requires NVIDIA's GPU hardware for operation.

	\item[Hardware-in-the-loop (HITL) Channel Emulators:] One of the first \gls*{hitl} channel-emulator enablers was proposed in an early work \cite{peter_emu}, which uses \gls*{fpga} to emulate wireless channels by applying a \gls*{fir} filter to the received signal.
	      Work in \cite{peter_emu} also uses a host PC (emulation controller) to control and emulate the mobility of connected nodes and change the \gls*{fir} filter taps accordingly on the \gls*{fpga}.
	      A decade later, \gls*{darpa} hosted the Spectrum Collaboration Challenge (SC2) \cite{darpa_sc2}, which led to the creation of the Colosseum \cite{colosseum} platform, the largest \gls*{hitl} channel emulator supporting up to 256 nodes with X310 \glspl*{usrp}.
	      In a recent research effort \cite{colosseum_hil} carried out by the Colosseum team, specific wireless technologies (Wi-Fi, \gls*{lte}, \gls*{5g}) are emulated with and without mobility on Colosseum Massive Channel Emulator~(MCHEM).
	      Colosseum has also been validated as a digital twin of the Arena over-the-air testbed in cellular-network and Wi-Fi-jamming scenarios~\cite{colosseum_dt}.
	      Unlike this RF/HITL approach, \gls*{achem} provides software-only I/Q-level emulation through \gls*{vusrp} interfaces, allowing unmodified \gls*{uhd}-based applications to run without SDR hardware or dedicated RF-emulation infrastructure.
	      MCHEM emulates the channel through \gls*{fir} filters, and mobility is provided through {\em pre-calculated} coefficients, which are stored in a dedicated \gls*{fpga}'s memory and applied to the received signals.
	      There are commercial solutions available for wireless channel emulation, such as PROPSIM~\cite{propsim} from Keysight Technologies, Vertex from Spirent~\cite{vertex_spirent}, and RFNest \cite{rfnest} from BlueHalo.
	      The main drawbacks of \gls*{hitl} channel emulators are cost, hardware-bound node and channel counts, and constrained development flexibility.
\end{description}

\section{Conclusion}
This work presents \gls*{achem}, an \emph{open-source} real-time digital twin framework for \gls*{usrp}-based wireless systems \cite{achem}.
By combining \gls*{vusrp} (radio emulation) and \gls*{chem} (I/Q-level channel emulation), \gls*{achem} enables unmodified \gls*{uhd}-based \gls*{sdr} applications to run in a fully software environment.
We validated the framework end-to-end against over-the-air experiments at the AERPAW Lake Wheeler testbed \cite{aerpaw} using LTE, LTE handover, and 5G stacks under the same trajectories and configurations.
Across these experiments, \gls*{achem} demonstrated end-to-end operation of LTE and 5G stacks, including successful mobility-driven LTE handover.
The testbed comparisons quantify SNR differences under the evaluated configurations and identify residual discrepancies associated with the selected propagation model and unmodeled physical effects.
Performance benchmarks characterize processing latency and scalability on server and laptop platforms, where laptop was able to handle nine node LTE deployment.
Together, these results establish \gls*{achem} as a practical path for repeatable development, debugging, and pre-deployment testing of SDR systems without requiring physical radio hardware.
To support reproducibility and further research, we have open-sourced the \gls*{achem} framework, including the validation experiments.
Additional implementation details, documentation, and updates are available at the ACHEM project website~\cite{achem_website}.

\section*{Acknowledgment}
We thank Samuel O'Brien for guidance on UHD software and National Instruments for making UHD open source.

\printbibliography

\newpage
\section*{Biographies}

\begin{IEEEbiography}[{\includegraphics[width=1in,height=1.25in,clip,keepaspectratio]{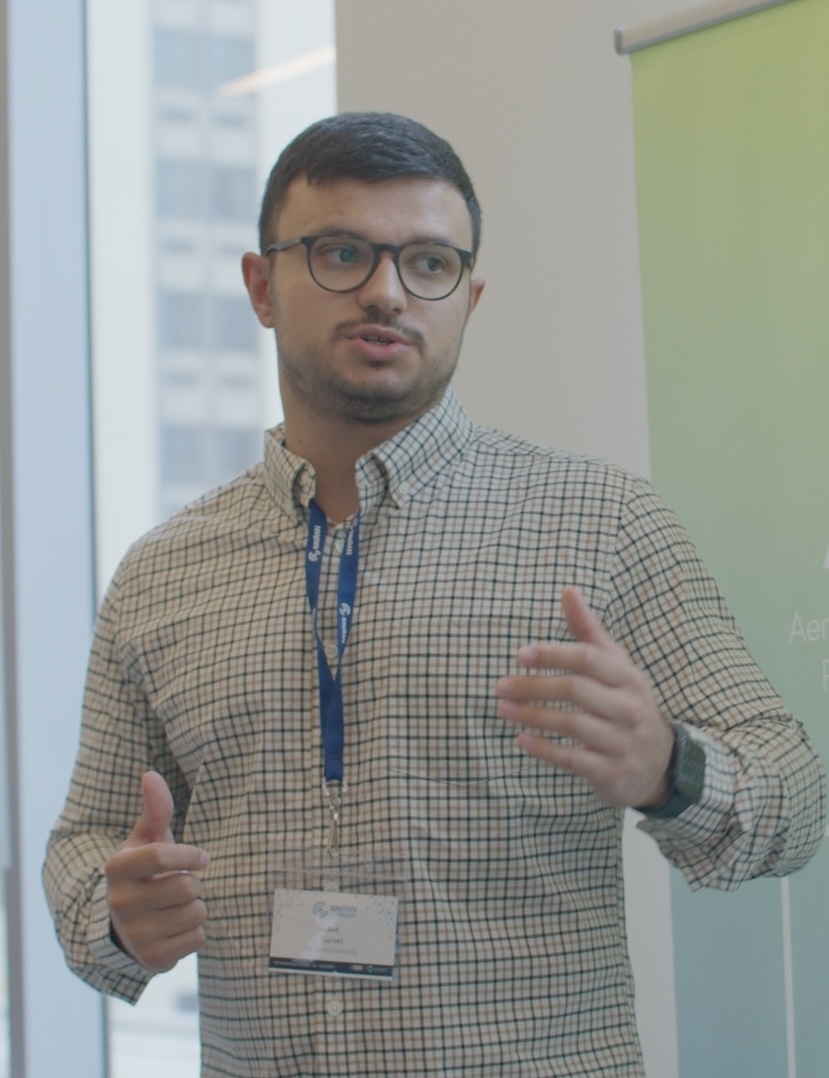}}]{An\i l G\"urses}
	An\i l G\"urses received the B.S. degree in Electrical and Electronics Engineering from \.{I}stanbul Medeniyet University. He is currently a PhD student in Electrical Engineering at NC State University. His research focuses on digital twins, wireless channel emulation, and UAV communications. He mainly works on software-defined radios and UAVs.
\end{IEEEbiography}

\begin{IEEEbiography}[{\includegraphics[width=1in,height=1.25in,clip,keepaspectratio]{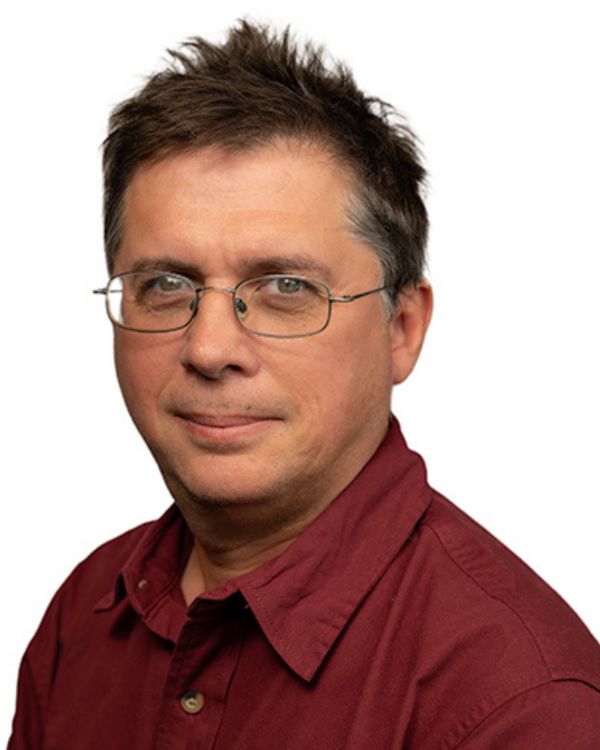}}]{Mihail L. Sichitiu}
	Mihail L. Sichitiu received the B.E. and M.S. degrees in Electrical Engineering from the Polytechnic University of Bucharest, in 1995 and 1996, respectively, and the Ph.D. degree in Electrical Engineering from the University of Notre Dame, Notre Dame, in 2001. He is currently a Professor with the Department of Electrical and Computer Engineering, NC State University. His research interests include wireless networking, wireless sensor networks, vehicular ad hoc networks, aerial wireless systems, and digital twins.
\end{IEEEbiography}

\vfill

\end{document}